\documentclass[twocolumn,showpacs,preprintnumbers,amsmath,amssymb,subfig]{revtex4-2}

\usepackage{graphicx}
\usepackage{hhline}

\begin{document}

\title{Discontinuous quantum and classical magnetic response of the pentakis dodecahedron \footnote{Dedicated to the memory of Professor Marshall Luban.}}

\author{N. P. Konstantinidis}
\affiliation{Department of Mathematics and Natural Sciences, The American University of Iraq, Sulaimani, Kirkuk Main Road, Sulaymaniyah, Kurdistan Region, Iraq}

\date{\today}

\begin{abstract}
The pentakis dodecahedron, the dual of the truncated icosahedron, consists of 60 edge-sharing triangles. It has 20 six-fold and 12 five-fold coordinated vertices, with the former forming a dodecahedron, and each of the latter connected to the vertices of one of the 12 pentagons of the dodecahedron. When spins mounted on the vertices of the pentakis dodecahedron interact according to the nearest-neighbor antiferromagnetic Heisenberg model, the two different vertex types necessitate the introduction of two exchange constants. As the relative strength of the two constants is varied the molecule interpolates between the dodecahedron and a molecule consisting only of quadrangles. The competition between the two exchange constants, frustration, and an external magnetic field results in a multitude of ground-state magnetization and susceptibility discontinuities. At the classical level the maximum is ten magnetization and one susceptibility discontinuities when the 12 five-fold vertices interact with the dodecahedron spins with approximately one-half the strength of their interaction. When the two interactions are approximately equal in strength the number of discontinuities is also maximized, with three of the magnetization and eight of the susceptibility. At the full quantum limit, where the magnitude of the spins equals $\frac{1}{2}$, there can be up to three ground-state magnetization jumps that have the $z$-component of the total spin changing by $\Delta S^z=2$, even though quantum fluctuations rarely allow discontinuities of the magnetization. The full quantum case also supports a $\Delta S^z=3$ discontinuity. Frustration also results in nonmagnetic states inside the singlet-triplet gap. These results make the pentakis dodecahedron the molecule with the largest number of magnetization and susceptibility discontinuities from the quantum to the classical level, taking its size also into account.
\end{abstract}


\maketitle

\section{Introduction}
\label{sec:introduction}

The antiferromagnetic Heisenberg model (AHM) is a prototype of strongly correlated electronic behavior, describing interactions between localized spins \cite{Auerbach98,Fazekas99}. Of special interest are lattices and molecules where in the classical lowest-energy configuration (LEC) antiferromagnetic interactions are not associated with parallel spins due to competing interactions, something known as frustration \cite{Lhuillier01,Misguich03}. Frustration originates in the topology of the lattice or molecule in question, which defines the geometry of spin interactions. It has important consequences both at the classical and quantum level including phases without conventional order such as the spin-liquid phase, singlet excitations inside the singlet-triplet gap, and magnetization plateaus in an external magnetic field added to the AHM \cite{NPK09,Ramirez05,Greedan01,Schnack10,Schnack19}.

Another important consequence of frustration are discontinuities in the LEC magnetization and susceptibility as a function of an external field.
This is unexpected as the AHM lacks any anisotropy in spin space, which is typically the origin of discontinuous magnetic behavior, as for example in the Ising model. Instead, the origin of the jumps is the frustrated connectivity of the interactions between spins. Ground-state magnetization discontinuities are a generic feature of fullerene molecules at the classical and the full quantum limit, as are discontinuities in the magnetic susceptibility at the classical level \cite{Coffey92,NPK05,NPK07,NPK16,NPK17,NPK18}. The icosahedron, which is a Platonic solid \cite{Plato}, is the smallest molecule associated with noncontinuous magnetic response \cite{Schroeder05,NPK15}. Magnetization discontinuities have also been found for extended systems \cite{Schulenburg02,Nakano13,Nakano15}. The addition of higher order isotropic exchange-interaction terms in the Hamiltonian enriches the discontinuous LEC magnetic response \cite{NPK16-1,NPK17-1}. Recently theories for the calculation of the ground states of classical Heisenberg spin systems have been developed \cite{Schmidt17,Schmidt17-1,Schmidt17-2,Schmidt17-3,Florek19,Schmidt20,Schmidt20-1}.

The fullerene molecules form a family very closely connected to noncontinuous magnetic response, as already mentioned. They are made of 12 pentagons and $\frac{n}{2}-10$ hexagons, where $n$ is the number of vertices of the molecule, which are three-fold coordinated \cite{Fowler95}. Their dual molecules consist of triangles, and have twelve five-fold coordinated vertices that originate in the fullerene pentagons, while the rest of the vertices originate in the fullerene hexagons and have six nearest neighbors. The total number of vertices of the dual fullerene molecules is $N=\frac{n}{2}+2$, while their spatial symmetry is the one of the molecule they are derived from.

The fullerene molecules associated with ground-state quantum magnetization discontinuities belong to the icosahedral $I_h$ spatial symmetry group \cite{NPK05,NPK07,NPK16}, which has the highest number of operations among point symmetry groups, equal to 120 \cite{Altmann94}. When placed in an increasing magnetic field the projection of the total spin $S$ along the field axis $S^z$, except from the standard changes between successive sectors with $\Delta S^z=1$, presents discontinuities with $\Delta S^z=2$ where the lowest-energy state of specific $S^z$ sectors never becomes the ground state in the field. For the smallest $I_h$ fullerene molecule, the dodecahedron ($n=20$), this occurs once for spins with magnitude $s=\frac{1}{2}$ and twice for $s=1$. For bigger $I_h$ fullerenes there is at least one discontinuity for $s=\frac{1}{2}$.

Molecular ground-state magnetization discontinuities in the absence of anisotropy are much more often found for classical spins, generating an increased interest to find molecules with jumps when the spins mounted at their vertices have $s=\frac{1}{2}$. The quantum discontinuities of the $I_h$-symmetry molecules provide the motivation to consider their equally high-symmetric duals. The icosahedron, the dual of the dodecahedron and the smallest $I_h$-symmetry fullerene dual, has $N=12$ equivalent vertices and no discontinuity for small $s$ \cite{Schroeder05,NPK15,Vaknin14,Engelhardt14,Hucht11,sahoo11,Sahoo12,Strecka15,Karlova16,Karlova17,Karlova17-1}. The next-larger $I_h$ fullerene is perhaps the most well-known among this family of molecules, the truncated icosahedron ($n=60$) \cite{Kroto87}, which has also been studied within the context of the AHM \cite{Coffey92,NPK07,Rausch20}. Its dual is the pentakis dodecahedron with $N=32$ (Fig. \ref{fig:pentakisdodecahedronclusterconnectivity}). It has 20 six-fold coordinated vertices that form a dodecahedron, and 12 five-fold coordinated vertices with each connected to the vertices of one of the 12 pentagon faces of the dodecahedron. The pentakis dodecahedron consists of 60 edge-sharing triangles, the smallest frustrated unit. Since the five-fold have only six-fold coordinated vertices as nearest neighbors, while the six-fold coordinated vertices also have nearest neighbors of their own kind, there are two geometrically different types of nearest-neighbor bonds, which are taken to be nonnegative and are associated with two different exchange interaction constants in the AHM.

In this paper, in order to investigate correlations between ground-state magnetic response and spatial symmetry for dual fullerene molecules of $I_h$ symmetry, and also the existence of classical and quantum magnetic response discontinuities, the AHM on the pentakis dodecahedron is considered. The ground-state magnetic response is calculated as a function of the relative strength of the two types of inequivalent nearest-neighbor bonds. This allows the calculation of the magnetic properties between two well-defined limits. When the exchange interaction between the five-fold and the six-fold coordinated vertices is zero, the AHM reduces to the one on the dodecahedron. At the opposite limit where only these bonds are nonzero, the AHM is the one of a molecule consisting only of quadrangles, lacking any frustration. Inbetween these two limits the competition between the two types of bonds, frustration, and an external magnetic field generate a ground-state magnetization response rich in discontinuities both for classical and $s=\frac{1}{2}$ spins.

The ratio of the exchange constants between six-fold and five-fold and between only six-fold coordinated spins is called $J$. For $J \lesssim 1$ the classical LEC magnetization discontinuities of $J=0$, which correspond to the dodecahedron \cite{NPK05,NPK07}, survive for higher magnetic fields (Fig. \ref{fig:lowestenergyconfigurationdiscontinuities}). For lower fields a series of discontinuities appears, which reach a maximum of ten of the magnetization and one of the susceptibility for $J \sim 0.5$. For $J \sim 1$ the higher-magnetic-field dodecahedron discontinuities disappear, and for a window extending up to $J \sim 1.1$ the discontinuities change their character. The most discontinuous response in this $J$-range includes three magnetization and eight susceptibility jumps. For $J \gtrsim 1.1$ the effect of frustration diminishes and eventually only a single susceptibility discontinuity survives. The large number of discontinuities for $J \lesssim 1.1$ is a dramatic change from the single magnetization discontinuity of the smallest member of the family, the icosahedron. The LEC magnetic response of the pentakis dodecahedron is the most discontinuous to have been found taking also the size of the molecule into account.

The ground-state magnetization response is also rich in magnetization discontinuities at the full quantum limit $s=\frac{1}{2}$, something which is also in contrast with the icosahedron which has no jumps. The discontinuity of the dodecahedron ($J=0$) \cite{NPK05,NPK07,NPK16} survives again for $J \lesssim 1$, similarly to the classical case. Further similarities with the classical case are the new discontinuities for lower magnetic fields. There is a number of them for $J \lesssim 1$, and they even involve the $S^z=0$ state, in which case the discontinuity leads directly to the $S^z=3$ sector, having a strength of $\Delta S^z=3$. In accordance with the classical case, discontinuities occur up to $J \sim 1.1$, while for higher $J$ values they disappear due to the dwindling frustration. The maximum number of discontinuities for a single $J$ is three, including the case $J=1$ where the two exchange constants are equal. Their nature resembles the one of the $I_h$-symmetry fullerenes, where the two lowest-energy states on the sides of the discontinuity are nondegenerate. However, for small $J$ states with multiple degeneracy are connected with a magnetization step of $\Delta S^z=2$. Frustration also results in nonmagnetic excitations inside the singlet-triplet gap when the two exchange constants are comparable in strength.




The plan of this paper is as follows: Sec. \ref{sec:model} introduces the AHM with an added magnetic field, and Sec. \ref{sec:classicalspins} discusses the LEC and magnetic response for classical spins. Sec. \ref{sec:spinsonehalf} presents the low-energy spectrum and magnetization response in the quantum-mechanical case. Finally Sec. \ref{sec:conclusions} presents the conclusions.

\section{Model}
\label{sec:model}
\begin{figure}
\includegraphics[width=2.8in,height=2.8in]{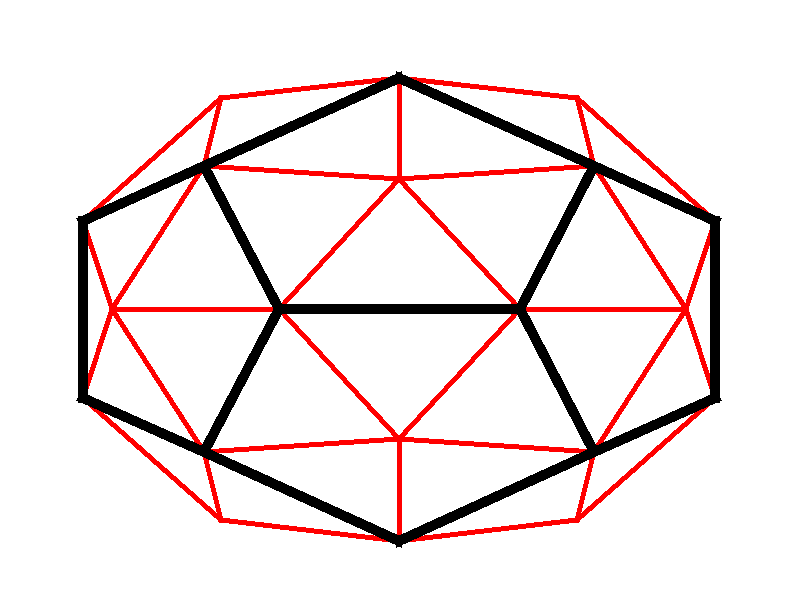}
\caption{(Color online) Top view of the pentakis dodecahedron projected on a plane. The (black) thick lines connect six-fold coordinated vertices, and the (red) thin lines six-fold and five-fold coordinated vertices. Each vertex has another one with the same coordination number placed directly below it, with the exception of the red vertices of the outer perimeter, and the leftmost and rightmost black vertices of the outer perimeter, which lie in the central plane of the molecule. The figure also gives the planar projection of the left-side and right-side view of the molecule. The spins $\vec{s}_i$ of Hamiltonian (\ref{eqn:Hamiltonian}) reside on the vertices and interact according to the lines.
}
\label{fig:pentakisdodecahedronclusterconnectivity}
\end{figure}

The pentakis dodecahedron has $N=32$ vertices (Fig. \ref{fig:pentakisdodecahedronclusterconnectivity}). It is the dual of the truncated icosahedron, and can be derived from the dodecahedron by introducing for each of its 12 pentagon faces a vertex that forms an edge with each of that pentagon's vertices. These twelve vertices form an icosahedron, and they are linked indirectly via vertices belonging to the dodecahedron. In Fig. \ref{fig:pentakisdodecahedronclusterconnectivity} the dodecahedron corresponds to the (black) thick lines, and its vertices are six-fold coordinated in the pentakis dodecahedron. The 12 extra spins are five-fold coordinated and generate the (red) thin lines, which link five-fold and six-fold coordinated vertices. The Hamiltonian of the AHM with an added external magnetic field term for spins $\vec{s}_i$ and $\vec{s}_j$ mounted on the vertices $i,j=1,\dots,N$ of the molecule is
\begin{eqnarray}
H & = & \sum_{<ij>} \textrm{} \vec{s}_i \cdot \vec{s}_j + J \sum_{<ij>^{'}} \textrm{} \vec{s}_i \cdot \vec{s}_j - h \sum_{i=1}^{N} s_i^z
\label{eqn:Hamiltonian}
\end{eqnarray}
The first two terms correspond to the (black) thick and the (red) thin lines of Fig. \ref{fig:pentakisdodecahedronclusterconnectivity} respectively, with the exchange constant of the first term defining the unit of energy, and the exchange constant $J$ of the second being nonnegative. The brackets in $<ij>$ and $<ij>^{'}$ indicate that interactions are limited to nearest neighbors. $h$ is the strength of an external magnetic field, taken to be directed along the $z$ axis. When $J=0$ the Hamiltonian reduces to the one of the dodecahedron and twelve uncoupled spins. When $J \to \infty$ the AHM reduces to the one of a molecule consisting of quadrangles linked together, and of the six nearest-neighbor bonds of each dodecahedron vertex only three are nonzero.

In Hamiltonian (\ref{eqn:Hamiltonian}) the minimization of the exchange energy competes with the one of the magnetic energy. Simultaneously the frustrated topology of the pentakis dodecahedron plays a very important role, as the frustration introduced by the pentagons at the $J=0$ limit competes with the frustration introduced by the triangles once $J \neq 0$. For classical spins minimization of the Hamiltonian gives the lowest-energy spin configuration as a function of $J$ and $h$ \cite{Coffey92,NPK05,NPK07,NPK13,NPK15,NPK15-1,NPK16,NPK16-1,NPK17,NPK18,NPK17-1,Machens13}. Each spin $\vec{s}_i$ is a classical unit vector with its direction defined by a polar $\theta_i$ and an azimuthal $\phi_i$ angle. A random initial direction is chosen for every spin and each angle is moved opposite to the direction of its gradient, until the energy minimum is reached. This procedure is repeated for different initial angles to ensure that the global energy minimum is found. Then the polar and azimuthal angles of the LEC and its symmetries are known, and its energy functional can be written.

For quantum-mechanical spins the Hamiltonian is block-diagonalized by taking into account $S^z$ and its spatial and spin symmetries \cite{Altmann94,NPK04,NPK05,NPK07,NPK09,NPK16,NPK15,NPK18,Machens13}. The point symmetry group is $I_h$, and the spin-inversion symmetry group is used when $S^z=0$. The degeneracy of each Hamiltonian subblock is then known. When the subblocks are small enough all eigenvalues and eigenstates are found. For larger matrices only Arnoldi \cite{Lehoucq96} and Lanczos diagonalization are possible. Comparison of the eigenenergies in different $S^z$ sectors allows characterization of the eigenstates also according to $S$, with each $S$ state being $2S+1$ times degenerate. The Hilbert space of the AHM comprises $2^{32}=4,294,967,296$ states, and the biggest $S^z$ subsector is $S^z=0$ with $601,080,390$ states. The biggest symmetry subsector is $H_g$ \cite{Altmann94} of $S^z=1$ with $23,585,037$ states, demonstrating the drastic dimensionality reduction in the biggest block-diagonalized Hamiltonian submatrix with the use of symmetries.

\section{Classical Spins}
\label{sec:classicalspins}

\subsection{Ground State in Zero Magnetic Field}
\label{subsec:groundstatezerofieldclassical}

\begin{figure}
\includegraphics[width=3.3in,height=2.8in]{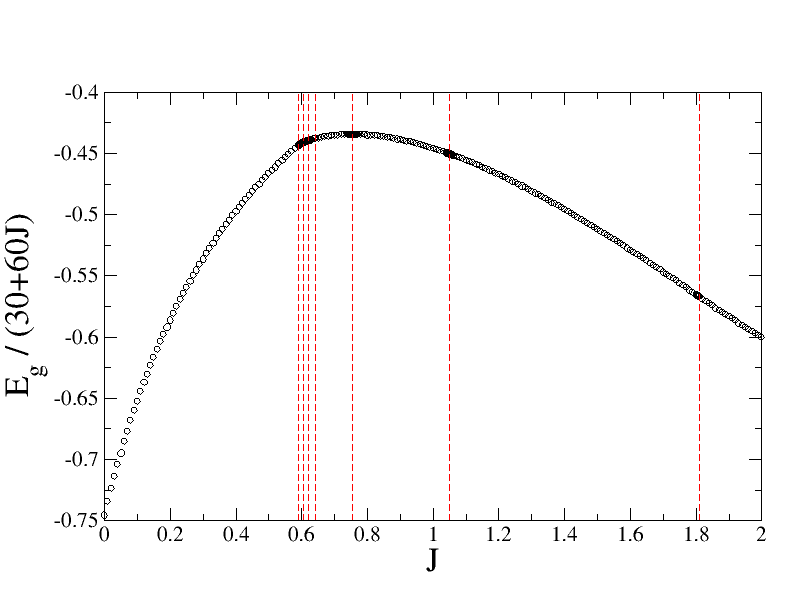}
\caption{(Color online) The (black) circles give the reduced energy of the classical zero-field ground state per bond $\frac{E_g}{30+60J}$ of Hamiltonian (\ref{eqn:Hamiltonian}) plotted as a function of $J$. The limiting value for $J \to \infty$ is -1. The (red) long-dashed vertical lines show the $J$ values where the symmetry of the LEC changes.
}
\label{fig:pentakisdodecahedronzerofieldenergy}
\end{figure}

First, the ground state of Hamiltonian (\ref{eqn:Hamiltonian}) is calculated for classical spins and zero magnetic field. When $J=0$ nearest-neighbor spins that reside on the vertices of the dodecahedron form a relative angle of cos$^{-1}(-\frac{\sqrt{5}}{3})$ \cite{Coffey92,NPK01,Schmidt03}. Once $J$ becomes nonzero, each nondodecahedron spin feels the influence of its five nearest neighbors, and its relative angle with each one of them equals cos$^{-1}(-\sqrt{\frac{5-2\sqrt{5}}{15}})$. The dodecahedron spins are not changing their relative direction with respect to the $J=0$ case, and the total ground state magnetization is zero. The relative angle between the nearest nondodecahedron spins equals cos$^{-1}(-\frac{\sqrt{5}}{5})$, equal to the angle between nearest-neighbors in the zero-field ground state of the icosahedron \cite{NPK15,Schroeder05,Schmidt03}. The ground state energy per bond $\frac{E_g}{30+60J}$ initially decreases with $J$ (Fig. \ref{fig:pentakisdodecahedronzerofieldenergy}), showing that frustration is getting stronger with increasing $J$.

\begin{figure}
\includegraphics[width=3.5in,height=2.5in]{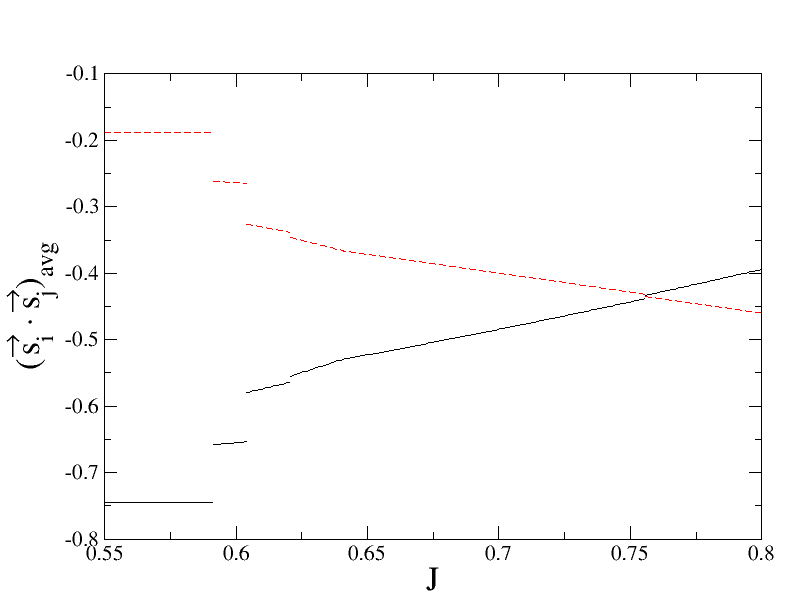}
\caption{(Color online) Average value of the nearest-neighbor correlations $\vec{s}_i \cdot \vec{s}_j$ as a function of $J$ in the classical zero-field LEC of Hamiltonian (\ref{eqn:Hamiltonian}). The (black) solid line is the average value for bonds between six-fold coordinated spins, and the (red) dashed line the average value for $J$ bonds between six-fold and five-fold coordinated spins.
}
\label{fig:pentakisdodecahedronzerofieldenergycontributionsbonds}
\end{figure}

\begin{figure}
\includegraphics[width=3.5in,height=2.5in]{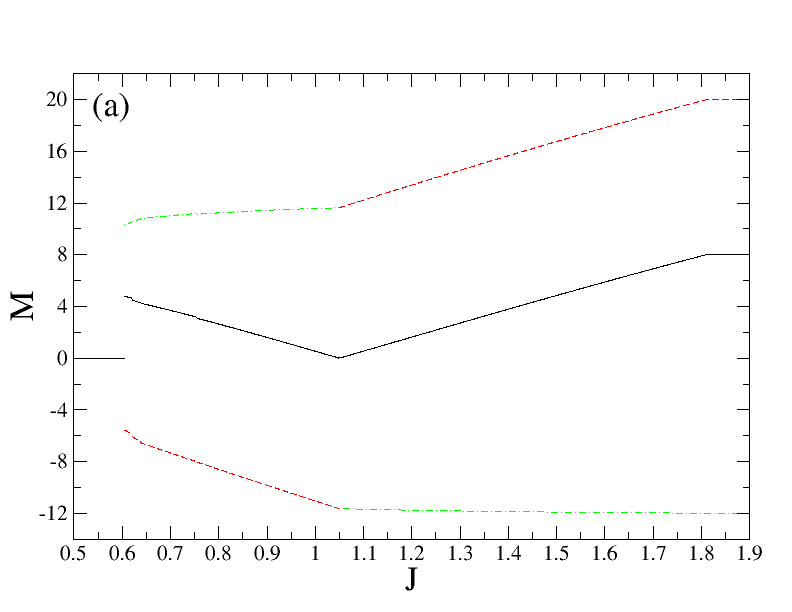}
\newline
\vspace{-15pt}
\includegraphics[width=3.5in,height=2.5in]{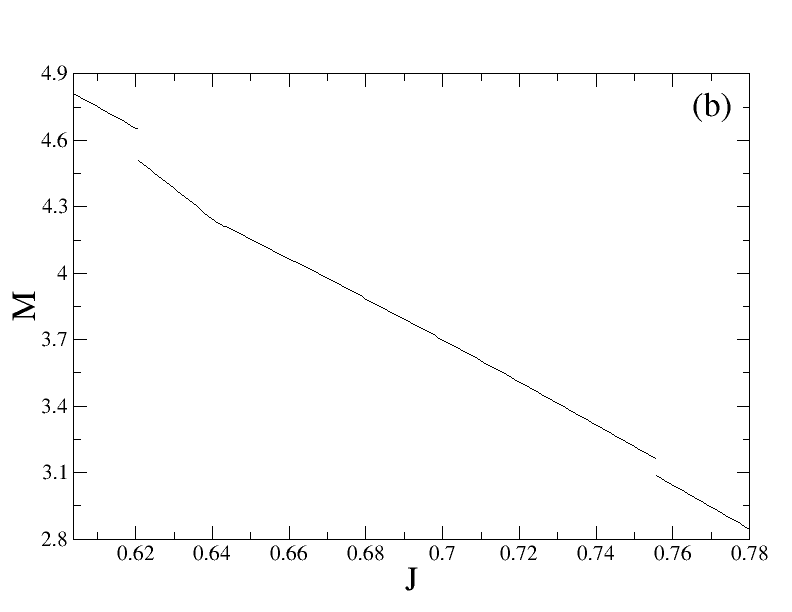}
\caption{(Color online) (a) Contributions along the direction of the classical zero-field lowest-energy total magnetization as a function of $J$. The (red) dashed line is the total contribution of the 20 six-fold coordinated spins, the (green) dot-dashed line the total contribution of the 12 five-fold coordinated spins, and the (black) solid line the total magnetization. (b) Part of (a) with the total magnetization shown in greater detail.
}
\label{fig:pentakisdodecahedronzerofieldmagnetizationcontributions}
\end{figure}

\begin{figure}
\includegraphics[width=3.5in,height=2.5in]{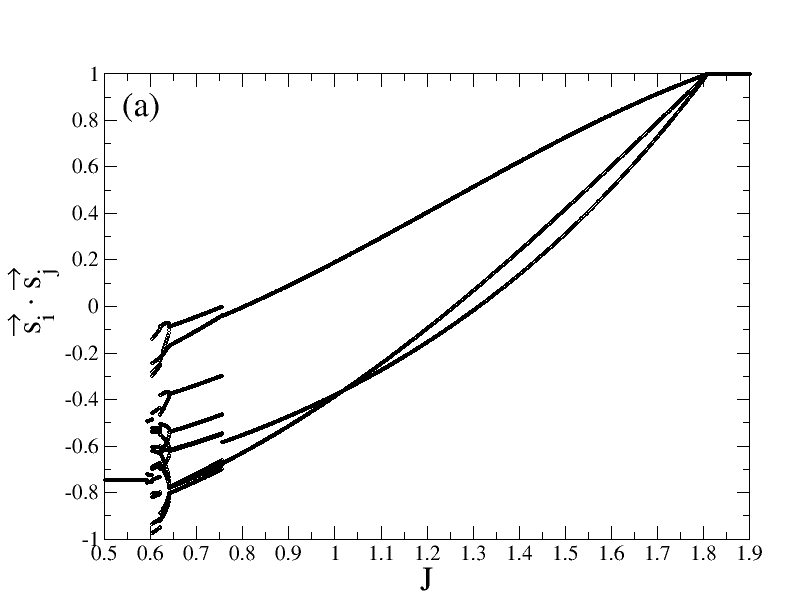}
\newline
\vspace{-15pt}
\includegraphics[width=3.5in,height=2.5in]{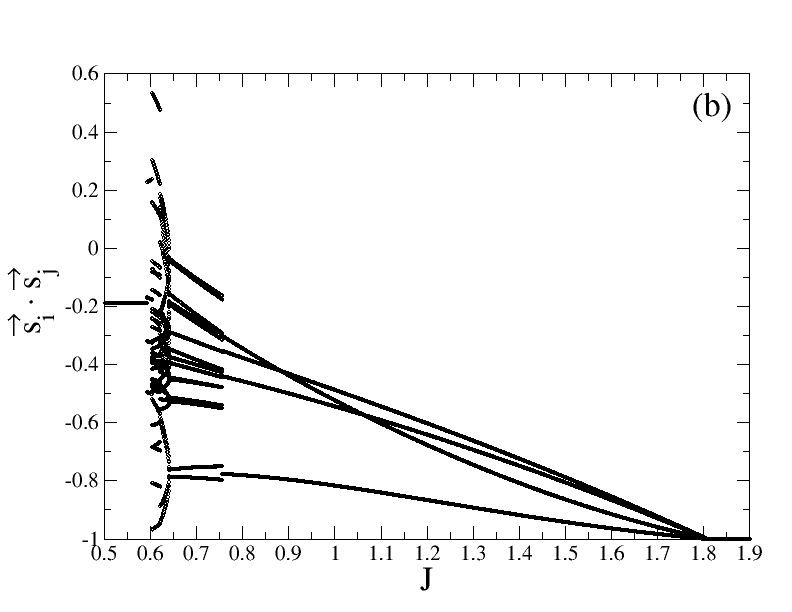}
\caption{Nearest-neighbor correlations $\vec{s}_i \cdot \vec{s}_j$ in the classical zero-field LEC of Hamiltonian (\ref{eqn:Hamiltonian}) as a function of $J$. (a) Correlations between six-fold coordinated spins, and (b) correlations between six-fold and five-fold coordinated spins.
}
\label{fig:pentakisdodecahedronzerofieldcorrelationscontributions}
\end{figure}

The LEC changes for $J \geq 0.591550$, with the spins reducing now the energy of the $J$ bonds more efficiently (Fig. \ref{fig:pentakisdodecahedronzerofieldenergycontributionsbonds}). The total magnetization remains zero (Fig. \ref{fig:pentakisdodecahedronzerofieldmagnetizationcontributions}), but the symmetry is reduced with the number of unique nearest-neighbor correlations increasing (Fig. \ref{fig:pentakisdodecahedronzerofieldcorrelationscontributions}). The proliferation of unique nearest-neighbor correlation functions continues for $J > 0.603929$, where the total ground state magnetization discontinuously acquires a finite value, with the 12 five-fold coordinated spins having a bigger net magnetization than the 20 dodecahedron spins.
Now the correlation functions associated with five-fold coordinated spins become comparable, at least in their minimum value, with the ones between spins residing on the dodecahedron.

The LEC changes again with a magnetization discontinuity at $J=0.620646$,
and then with a susceptibility discontinuity at $J=0.64075$ (Fig. \ref{fig:pentakisdodecahedronzerofieldmagnetizationcontributions}). The average nearest-neighbor correlation of the five-fold coordinated spins keeps strengthening its antiferromagnetic character for stronger $J$, at the expense of the energy of the dodecahedron spins (Fig. \ref{fig:pentakisdodecahedronzerofieldenergycontributionsbonds}).
The next change of the LEC occurs at $J=0.755654$ with a discontinuity in the total magnetization.
This value of $J$ marks the point of strongest frustration, with the ground state energy per bond $\frac{E_g}{30+60J}$ becoming maximum and equal to -0.43441574 (Fig. \ref{fig:pentakisdodecahedronzerofieldenergy}), slightly higher than $-\frac{\sqrt{5}}{5}$, the zero-field ground-state energy per bond of the AHM on the icosahedron \cite{NPK15,Schroeder05,Schmidt03}. Exactly at this point the average nearest-neighbor correlation between five-fold and six-fold coordinated spins becomes more antiferromagnetic in character than the one between six-fold coordinated spins (Fig. \ref{fig:pentakisdodecahedronzerofieldenergycontributionsbonds}), and the ground-state symmetry increases (Fig. \ref{fig:pentakisdodecahedronzerofieldcorrelationscontributions}).

At $J=1.049684$ the symmetry of the ground state does not change, but the total magnetization, which had been decreasing since it acquired a finite value, becomes zero and then starts to increase again for higher $J$ (Fig. \ref{fig:pentakisdodecahedronzerofieldmagnetizationcontributions}). Simultaneously, the total magnetization of the 20 six-fold coordinated spins becomes bigger than the one of the 12 five-fold coordinated spins for the first time.
Starting at $J=\frac{5+\sqrt{5}}{4}$ (App. \ref{appendix:saturationJzeromagneticfield}) the five-fold and six-fold coordinated spins point in opposite directions (Fig. \ref{fig:pentakisdodecahedronzerofieldcorrelationscontributions}), and the ground state energy equals $30-60J$.

\subsection{Ground-State Magnetization Response in an External Field}
\label{subsec:groundstatemagnetizationresponseinanexternalfieldclassical}

\begin{figure*}
\centerline{
\includegraphics[width=0.55\textwidth,height=2.8in]{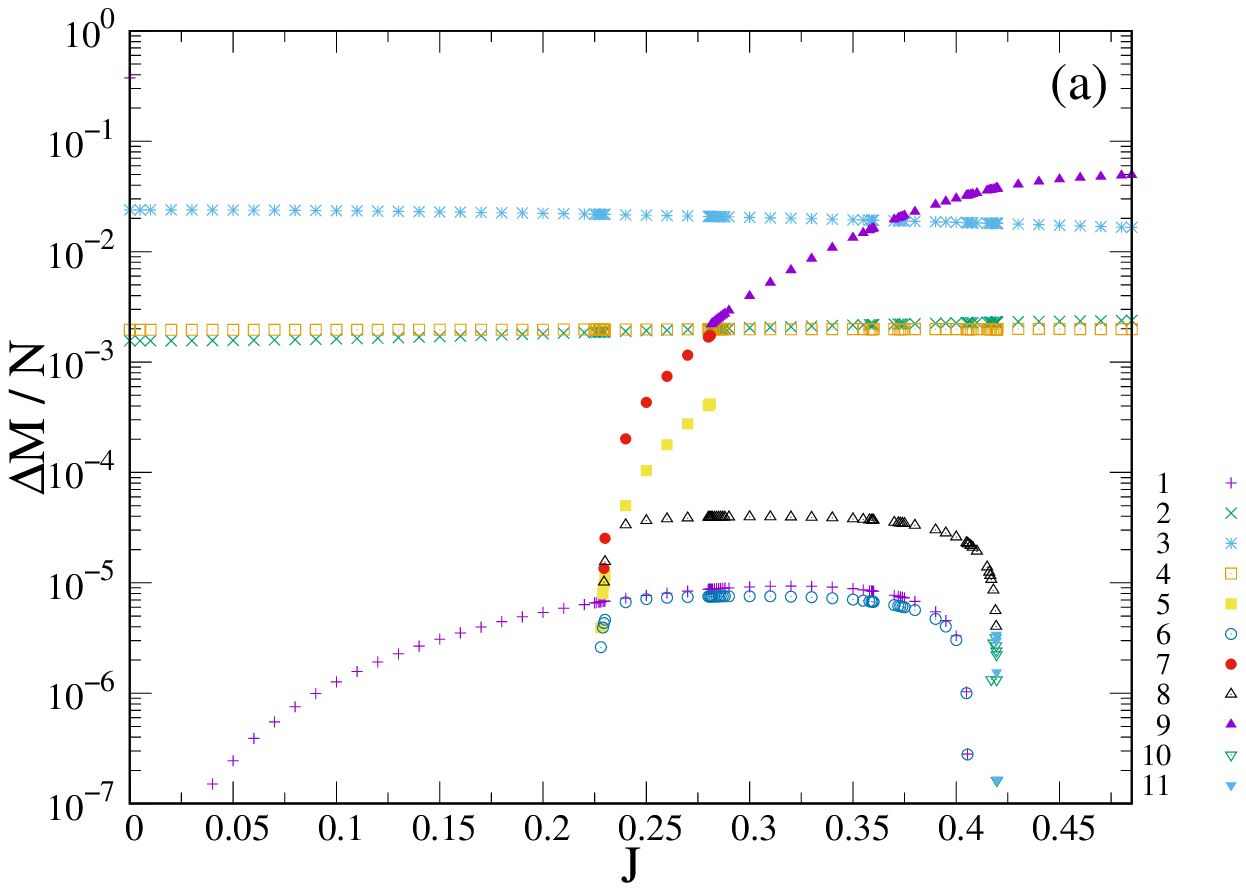}
\hspace{4pt}
\includegraphics[width=0.55\textwidth,height=2.8in]{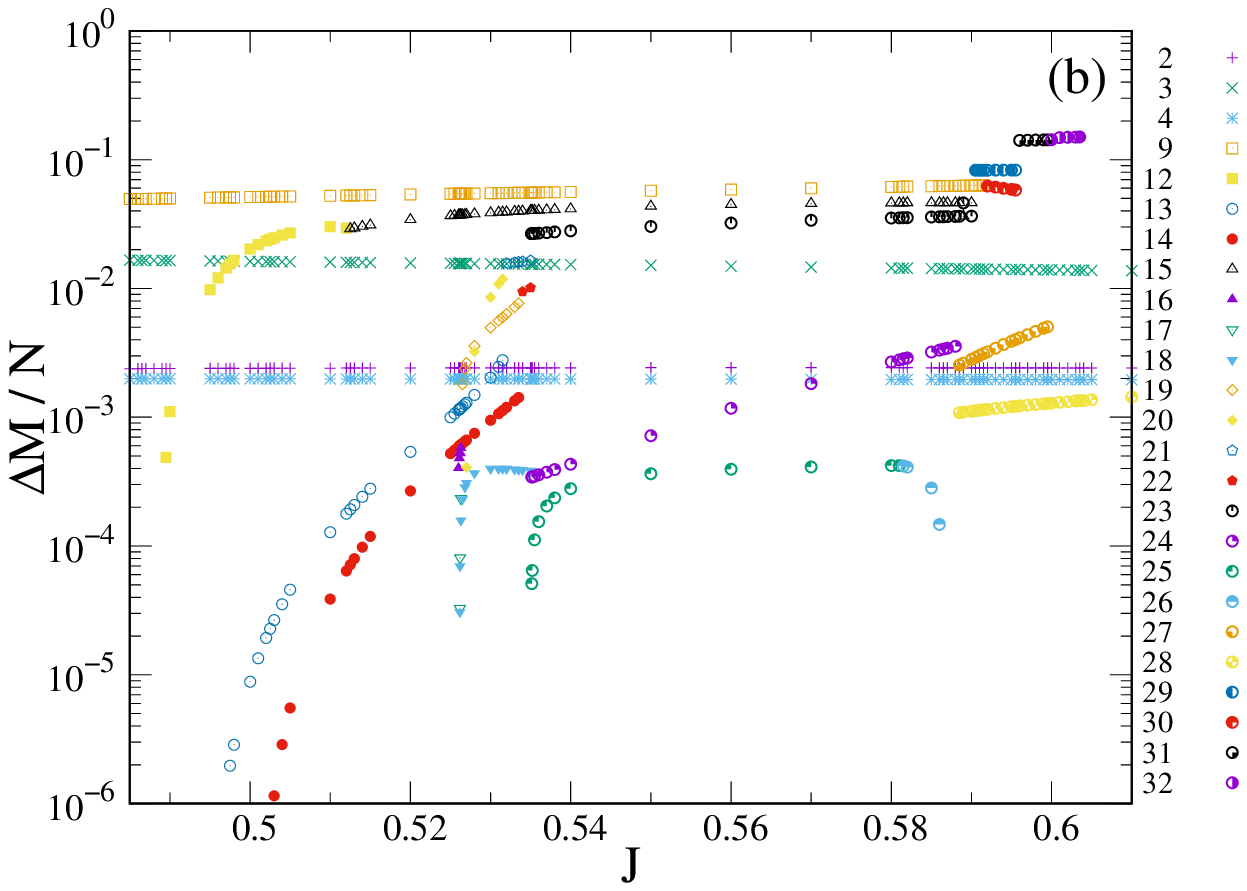}
}
\vspace{4pt}
\centerline{
\includegraphics[width=0.55\textwidth,height=2.8in]{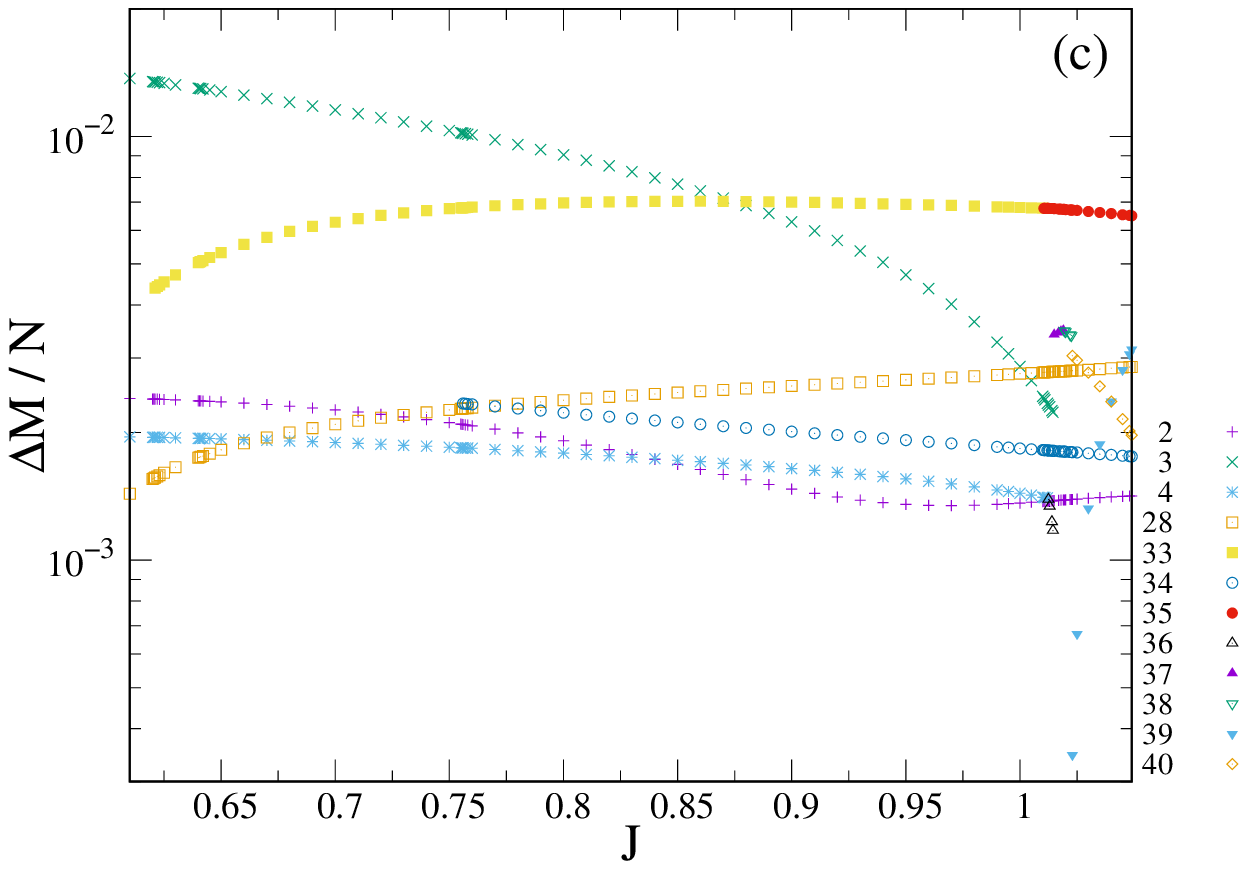}
\hspace{4pt}
\includegraphics[width=0.55\textwidth,height=2.8in]{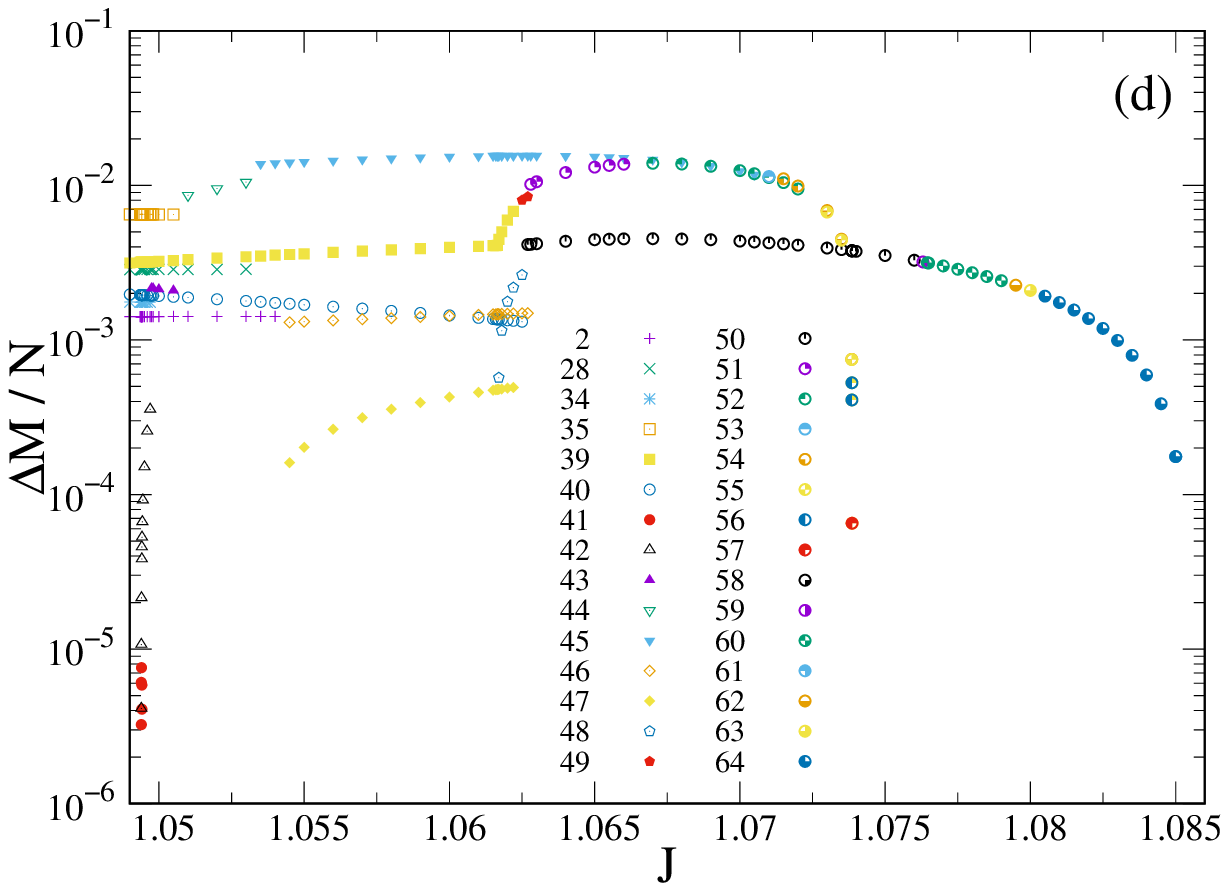}
}
\caption{(Color online) Magnitude of the magnetization discontinuities over the number of spins $\frac{\Delta M}{N}$ in the classical LEC of Hamiltonian (\ref{eqn:Hamiltonian}) as a function of $J$. The numbers correspond to the discontinuities in Table \ref{table:classicalmagnsuscdisc}, and increase with $J$.
}
\label{fig:lowestenergyconfigurationinaccessiblemagnetizations}
\end{figure*}

Once the magnetic field in Hamiltonian (\ref{eqn:Hamiltonian}) becomes nonzero, it competes with the exchange interactions for the minimization of the energy. The magnetic energy is most efficiently minimized when the spins point in the direction of the field. On the other hand, antiferromagnetic exchange supports interacting spins pointing in opposite directions. This competition results in a multitude of magnetization and susceptibility discontinuities in the LEC as the field is varied. The discontinuity fields are plotted in Figs \ref{fig:lowestenergyconfigurationdiscontinuities} and \ref{fig:lowestenergyconfigurationdiscontinuitiesfocus} as a function of $J$, and information on the discontinuities is given in Table \ref{table:classicalmagnsuscdisc}.
The strength of the magnetization discontinuities is plotted in Fig. \ref{fig:lowestenergyconfigurationinaccessiblemagnetizations}.
The saturation field $h_{sat}=\frac{1}{2}[3+\sqrt{5}+8J+\sqrt{(3+\sqrt{5})^2-4(3+\sqrt{5})J+8(3-\sqrt{5})J^2}]$ for $J \leq \frac{3}{20}(5+\sqrt{5})$, and $8J$ otherwise (App. \ref{appendix:saturationmagneticfield}).

When $J=0$ Hamiltonian (\ref{eqn:Hamiltonian}) includes the AHM on the dodecahedron, whose magnetic response has three magnetization discontinuities \cite{NPK05}. In addition, it includes twelve uncoupled spins which are polarized by an infinitesimal magnetic field, resulting in a fourth magnetization discontinuity at zero field. Looking at Fig. \ref{fig:lowestenergyconfigurationdiscontinuities} the higher-field discontinuities, which are associated with the dodecahedron, survive up to $J \sim 1$, showing that in this region the dodecahedron exchange prevails to the $J$ exchange. The dodecahedron correlations have been already shown to prevail for a significant $J$ range in Sec. \ref{subsec:groundstatezerofieldclassical}. Simultaneously, the dodecahedron is more resistant to an external magnetic field, having a higher saturation field than the triangle, whose type of interactions is introduced by the $J$ bonds.
 New discontinuities are generated for lower fields, related to the weaker exchange constant $J$ which competes with the field when it is relatively weak. 
For $J$ slightly higher than 1 the high-field dodecahedron discontinuities disappear and new discontinuities emerge, as $J$ becomes stronger. For even higher $J$ only two susceptibility jumps remain, and for $J > \frac{5+\sqrt{5}}{4}$ only one of them survives.

The maximum number of ground-state discontinuities occurs at $J = 0.526$, with ten of the magnetization and one of the susceptibility, and for $J$ between $1.0768$ and $1.0792$, with three and eight respectively (Table \ref{table:classicalmagnsuscdisc}). At $J=1$, where the two exchange constants are equal, there are six magnetization and three susceptibility jumps.
These discontinuity numbers are higher than the ones of the fullerene molecules \cite{Coffey92,NPK05,NPK07,NPK16,NPK17,NPK18}, also taking the number of vertices of the pentakis dodecahedron into account. The strongest magnetization jumps, except from the one at $J=0$ and $h=0+$, occur for the low-field discontinuities 31 and 32 (Table \ref{table:classicalmagnsuscdisc}), with a strength of approximately 15\% of the saturation magnetization (Figs \ref{fig:lowestenergyconfigurationdiscontinuitiesfocus}(f) and \ref{fig:lowestenergyconfigurationinaccessiblemagnetizations}(b)).

\begin{figure}
\includegraphics[width=3.5in,height=2.5in]{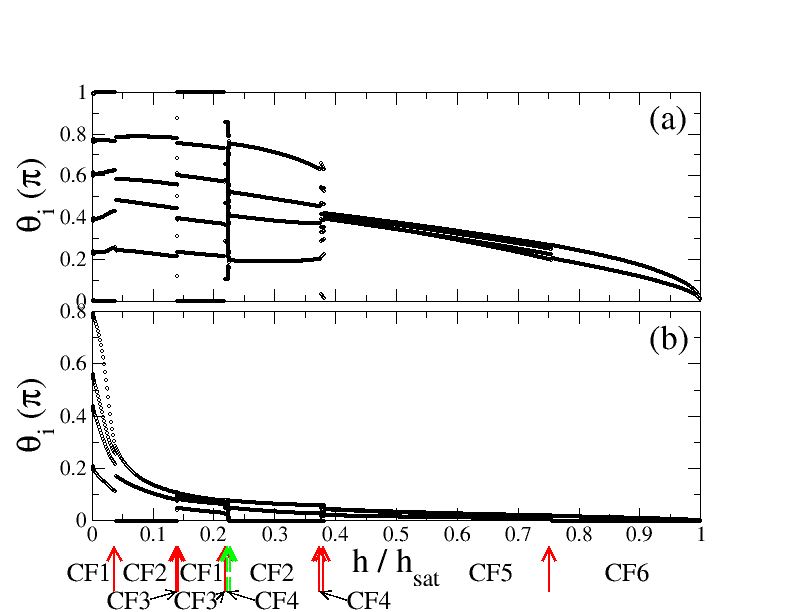}
\caption{(Color online) Classical LEC unique polar angles $\theta_i$ as a function of the magnetic field $h$ over its saturation value $h_{sat}$ for $J=0.3$ for (a) the six-fold coordinated spins, and (b) the five-fold coordinated spins. The (red) solid arrows point to the magnetization and the (green) dashed arrows to the susceptibility discontinuities. CF$_i$, $i=1,\dots,6$ refers to the lowest configurations for the different field ranges (Fig. \ref{fig:lowestenergyconfigurations}).
}
\label{fig:lowestenergyconfigurationanglesJ=0.3}
\end{figure}

\begin{figure}
\centerline{
\includegraphics[width=0.21\textwidth,height=1.2in]{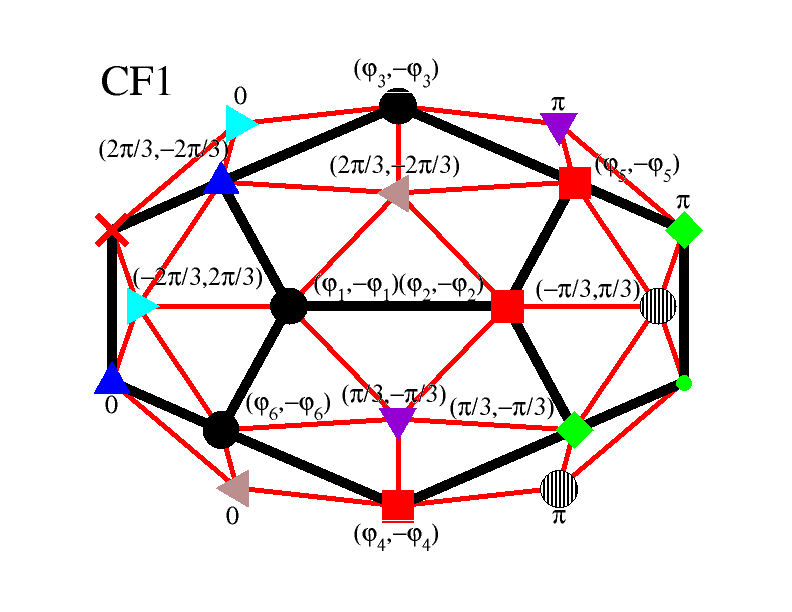}
\hspace{-15pt}
\includegraphics[width=0.21\textwidth,height=1.2in]{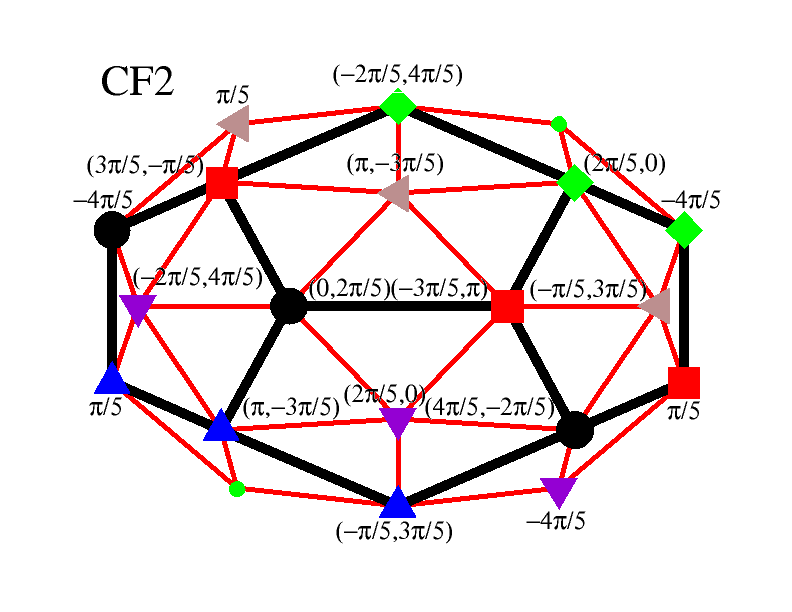}
\hspace{-15pt}
\includegraphics[width=0.21\textwidth,height=1.2in]{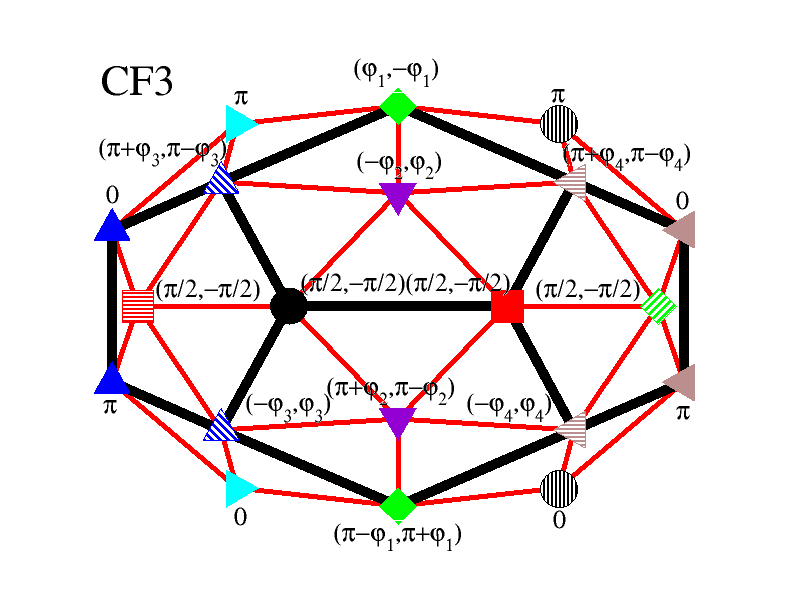}
}
\vspace{-10pt}
\centerline{
\includegraphics[width=0.21\textwidth,height=1.2in]{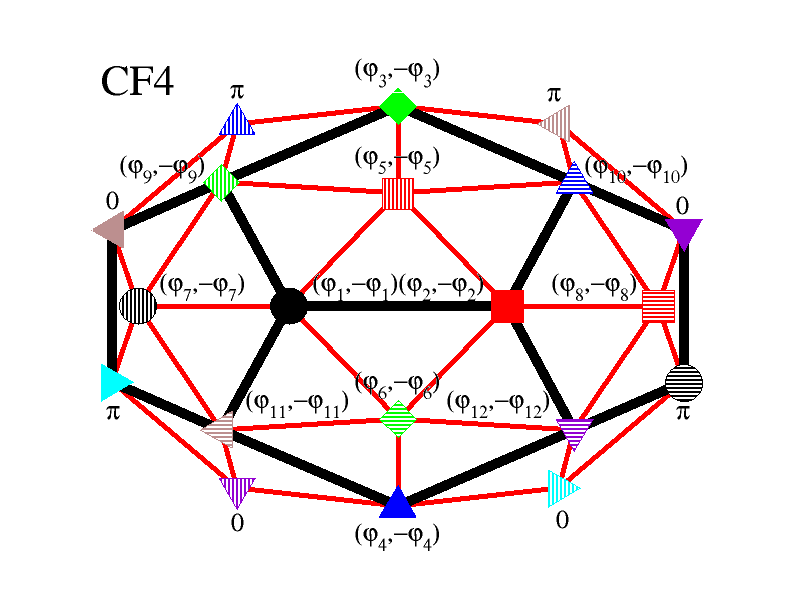}
\hspace{-15pt}
\includegraphics[width=0.21\textwidth,height=1.2in]{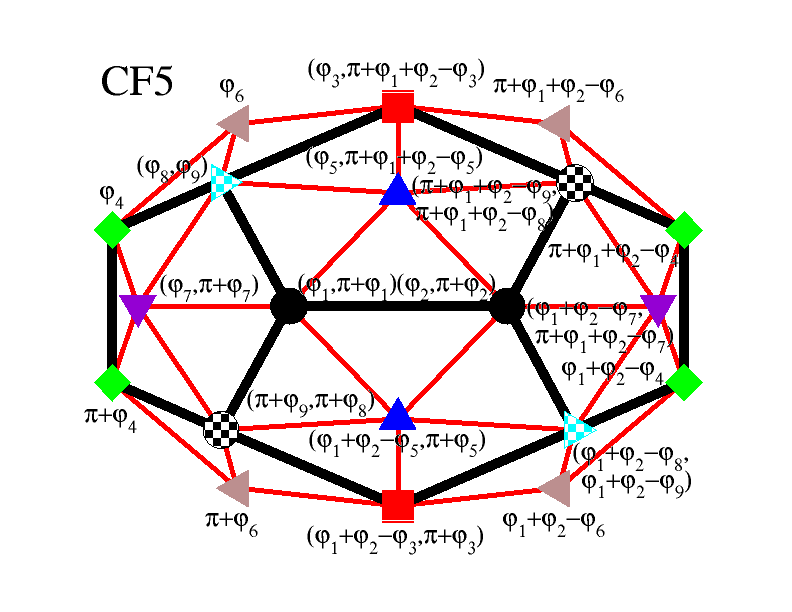}
\hspace{-15pt}
\includegraphics[width=0.21\textwidth,height=1.2in]{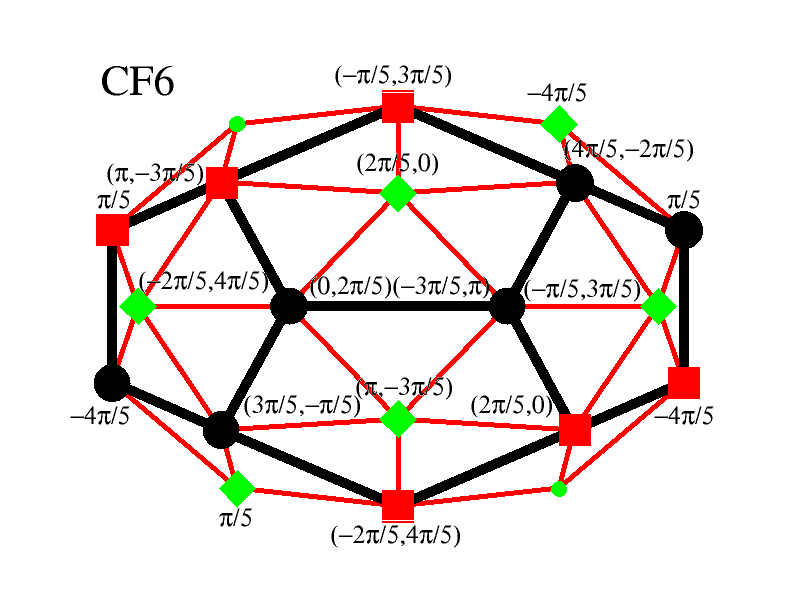}
}
\vspace{-10pt}
\centerline{
\includegraphics[width=0.21\textwidth,height=1.2in]{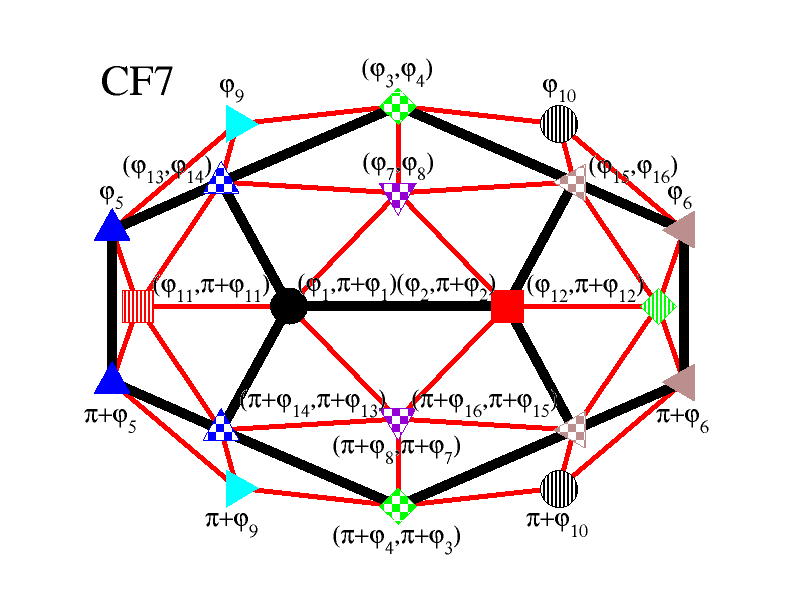}
\hspace{-15pt}
\includegraphics[width=0.21\textwidth,height=1.2in]{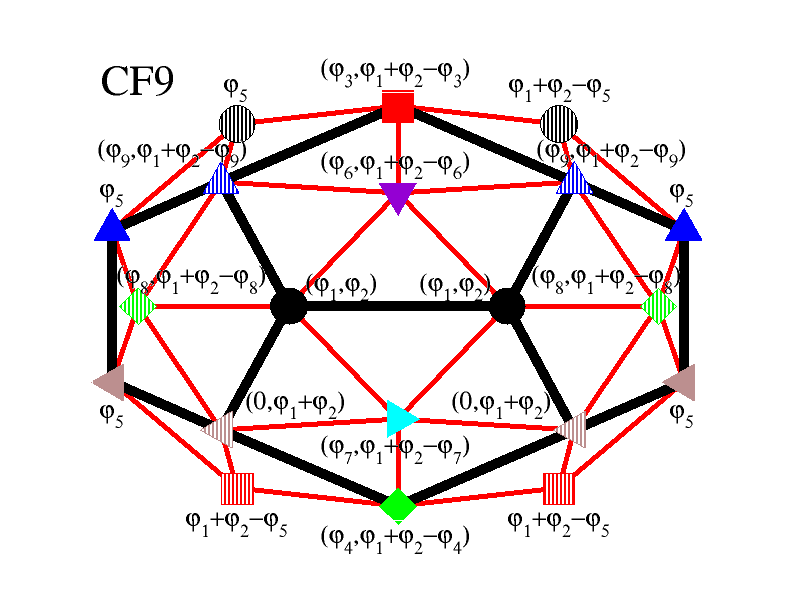}
\hspace{-15pt}
\includegraphics[width=0.21\textwidth,height=1.2in]{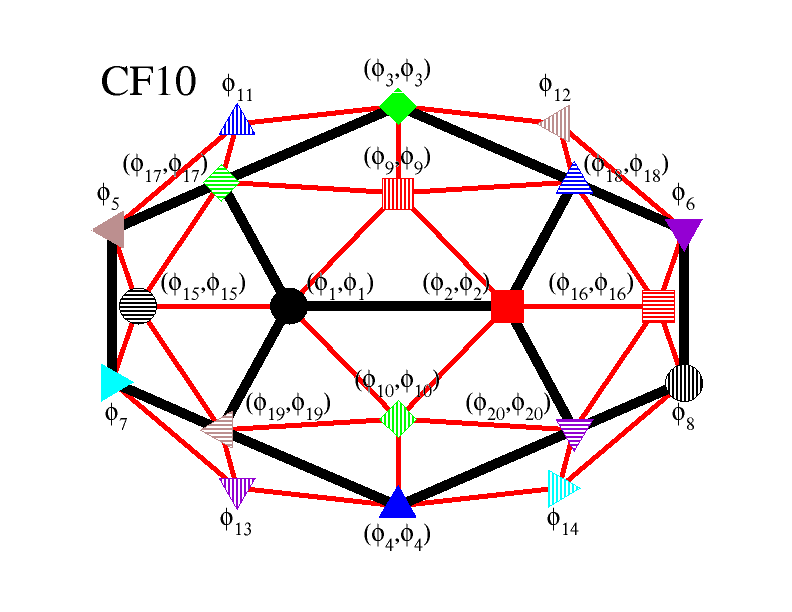}
}
\vspace{-10pt}
\centerline{
\includegraphics[width=0.21\textwidth,height=1.2in]{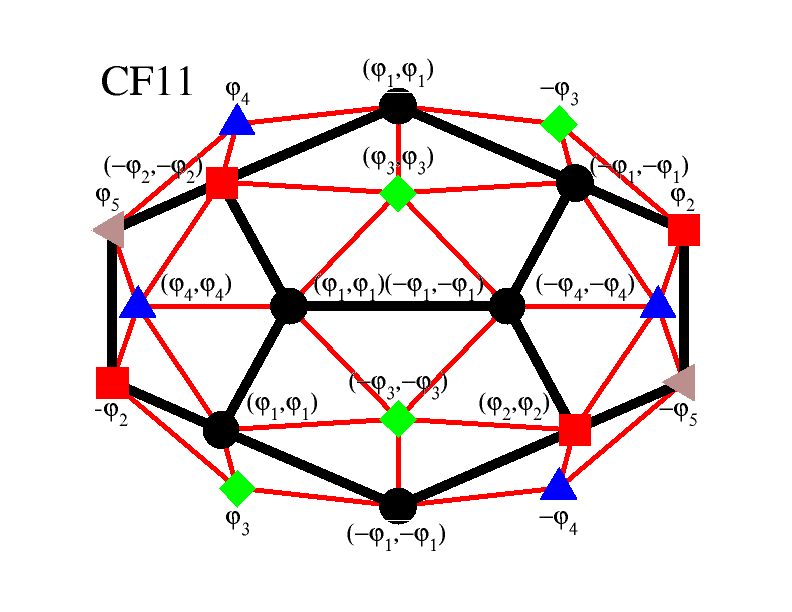}
\hspace{-15pt}
\includegraphics[width=0.21\textwidth,height=1.2in]{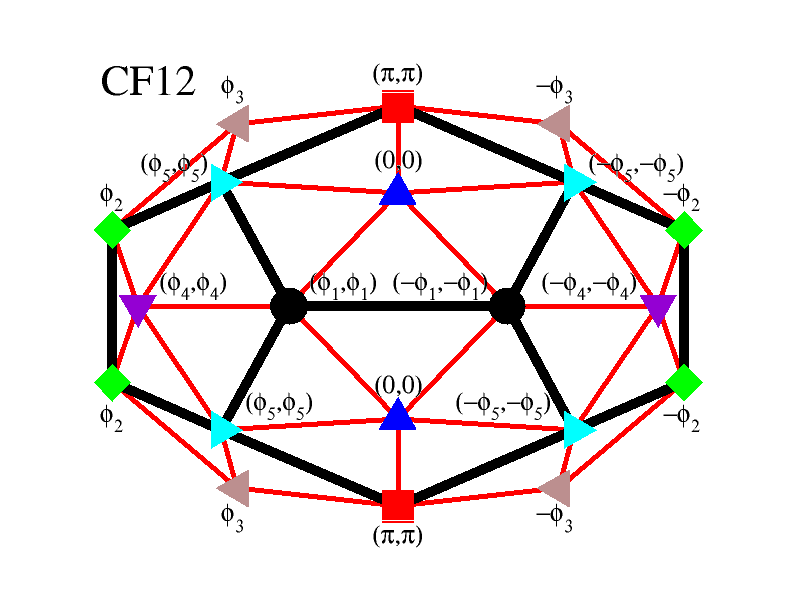}
\hspace{-15pt}
\includegraphics[width=0.21\textwidth,height=1.2in]{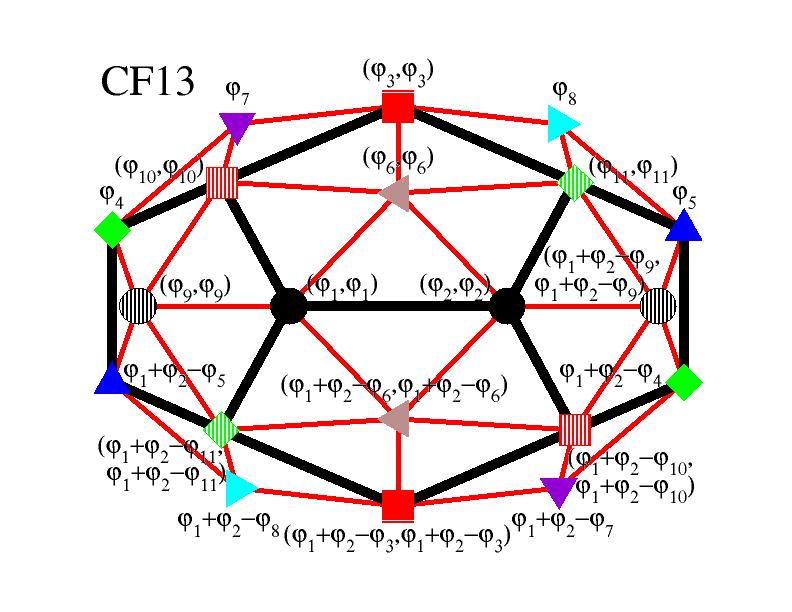}
}
\vspace{-10pt}
\centerline{
\includegraphics[width=0.21\textwidth,height=1.2in]{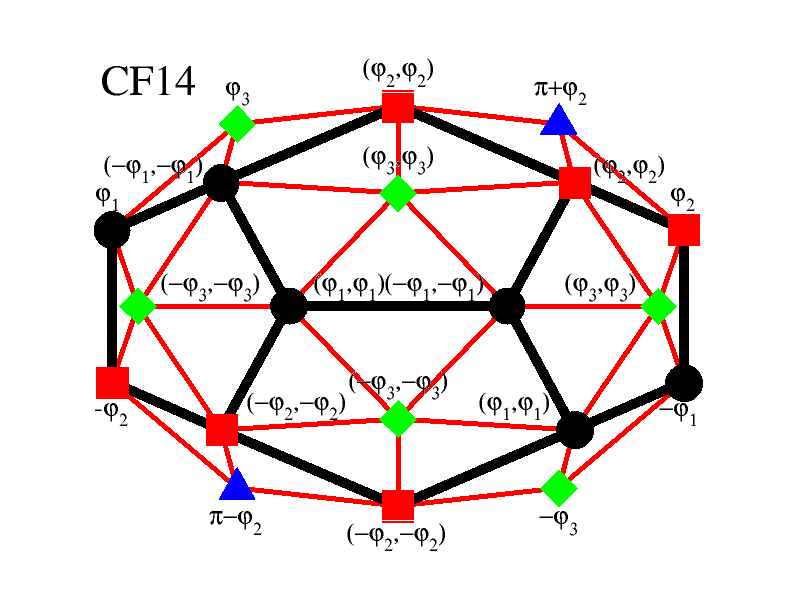}
\hspace{-15pt}
\includegraphics[width=0.21\textwidth,height=1.2in]{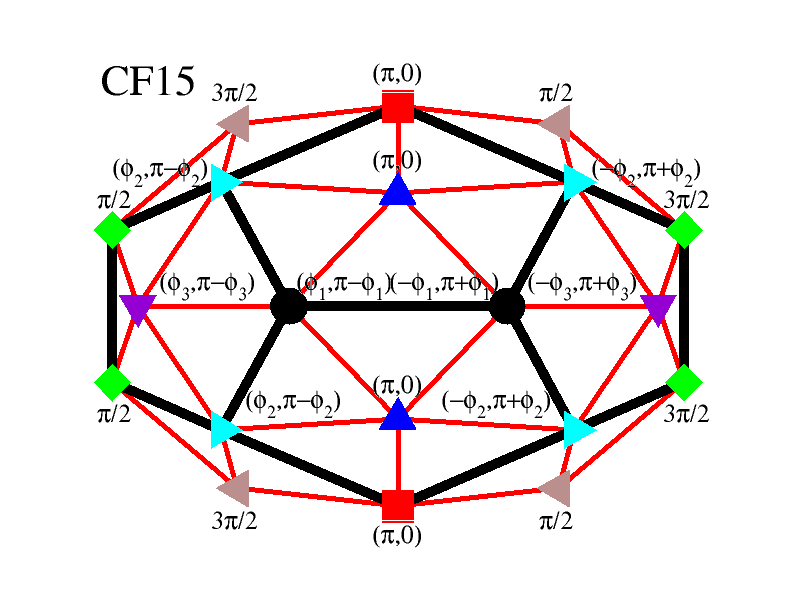}
\hspace{-15pt}
\includegraphics[width=0.21\textwidth,height=1.2in]{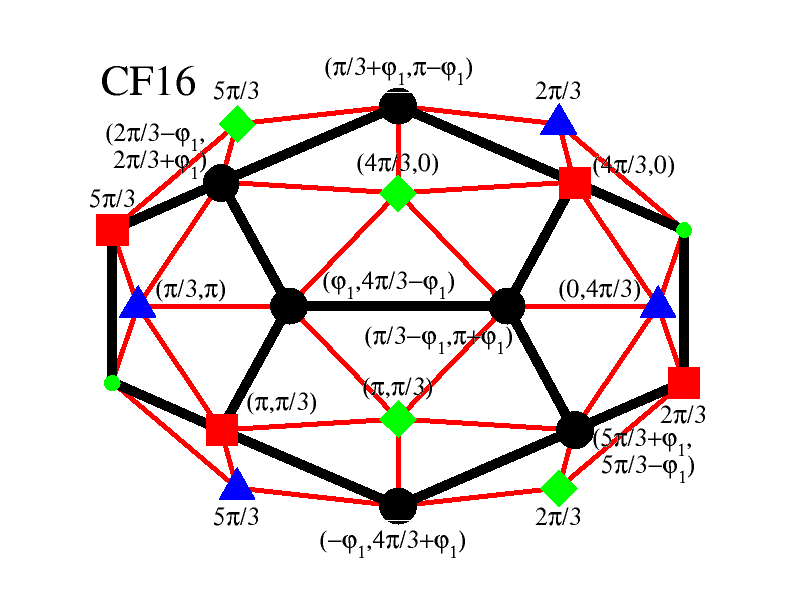}
}
\caption{(Color online) All the classical LECs CF$i$, $i=1,\dots,16$, of Hamiltonian (\ref{eqn:Hamiltonian}) for $J=0.3$, 1, and 1.08, for any $h$. Spins with the same polar angle are represented by the same symbol and fill pattern (and color). The spins right below them (Fig. \ref{fig:pentakisdodecahedronclusterconnectivity}) correspond to the same polar angle, with the exception of configurations CF5 anf CF7, where symbols with the checkerboard fill pattern exchange the polar angles of their lower spins. The small (green) circle corresponds to a spin along the field, and the (red) x to a spin antiparallel to the field. The azimuthal angles are explicitly written. When two numbers are given the first is for the spin above and the second for the spin directly below it. Configuration CF8 is very similar to configuration CF6, the only difference being that the two icosahedron spins are antiparallel rather than parallel to the field.
}
\label{fig:lowestenergyconfigurations}
\end{figure}

Fig. \ref{fig:lowestenergyconfigurationanglesJ=0.3} shows the dependence of the unique LEC polar angles on the magnetic field for $J=0.3$.
The polar angles corresponding to the five-fold coordinated spins rapidly decrease with the field due to the weakness of $J$.
On the other hand, the dodecahedron spins are more resistant to the influence of the magnetic field. The LECs corresponding to the different magnetic fields are shown in Fig. \ref{fig:lowestenergyconfigurations}.

\begin{figure}
\includegraphics[width=3.5in,height=2.5in]{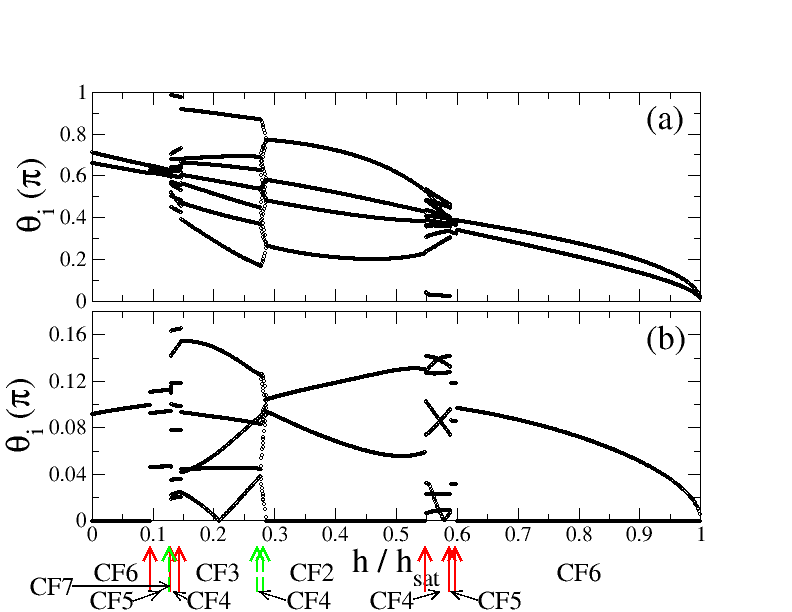}
\caption{(Color online) Classical LEC unique polar angles $\theta_i$ as a function of the magnetic field $h$ over its saturation value $h_{sat}$ for $J=1$ for (a) the six-fold coordinated spins, and (b) the five-fold coordinated spins. The (red) solid arrows point to the magnetization and the (green) dashed arrows to the susceptibility discontinuities. CF$_i$, $i=2,\dots,7$ refers to the lowest configurations for the different field ranges (Fig. \ref{fig:lowestenergyconfigurations}).
}
\label{fig:lowestenergyconfigurationanglesJ=1}
\end{figure}

\begin{figure}
\includegraphics[width=3.5in,height=2.5in]{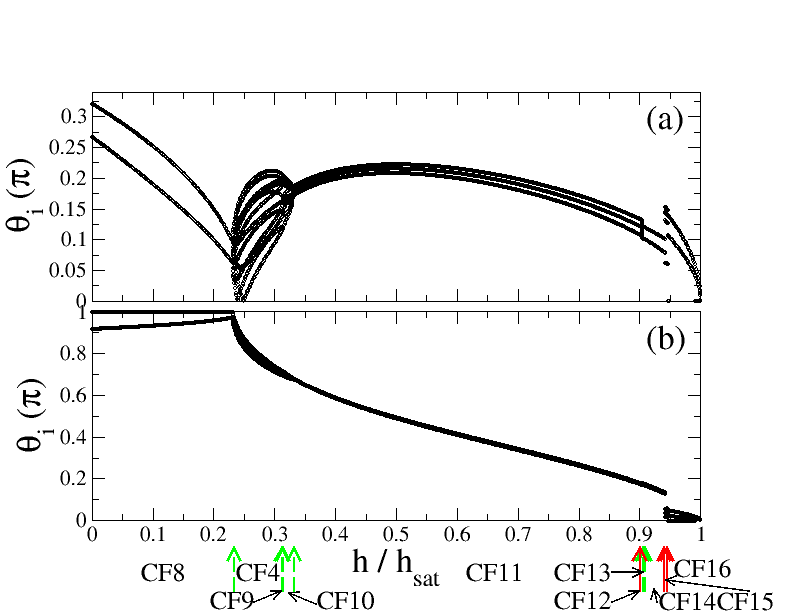}
\caption{(Color online) Classical LEC unique polar angles $\theta_i$ as a function of the magnetic field $h$ over its saturation value $h_{sat}$ for $J=1.08$ for (a) the six-fold coordinated spins, and (b) the five-fold coordinated spins. The (red) solid arrows point to the magnetization and the (green) dashed arrows to the susceptibility discontinuities. CF$_i$, $i=4$, and $i=8,\dots,16$ refers to the lowest configurations for the different field ranges (Fig. \ref{fig:lowestenergyconfigurations}).
}
\label{fig:lowestenergyconfigurationanglesJ=1.08}
\end{figure}

Fig. \ref{fig:lowestenergyconfigurationanglesJ=1} shows the evolution of the unique polar angles with the magnetic field for $J=1$. Now the five-fold coordinated spins have a higher magnetic energy for small fields, with the only new LEC introduced with respect to $J=0.3$ being CF7 (Fig. \ref{fig:lowestenergyconfigurations}). On the contrary, for a slightly higher $J=1.08$ many new LECs are introduced (Figs \ref{fig:lowestenergyconfigurationanglesJ=1.08} and \ref{fig:lowestenergyconfigurations}). Now the magnetic field has to work harder against the five-fold coordinated than the dodecahedron spins.

\section{Quantum Spins $s=\frac{1}{2}$}
\label{sec:spinsonehalf}

\subsection{Ground State in Zero Magnetic Field}
\label{subsec:groundstatezerofieldspinonehalf}

For quantum spins $s=\frac{1}{2}$ the ground state of Hamiltonian (\ref{eqn:Hamiltonian}) for zero magnetic field is listed in Table \ref{table:spinonehalfzerofieldgroundstate} as a function of $J$. For $J=0$ the 12 five-fold coordinated spins are not interacting with their nearest-neighbors, whose ground state is the one of the isolated dodecahedron which is a singlet \cite{NPK05}, making the ground state of the pentakis dodecahedron $2^{12}$ times degenerate. The ground state of the isolated dodecahedron belongs to the singly-degenerate $A_u$ irreducible representation of the $I_h$ symmetry group \cite{Altmann94}, and is symmetric with respect to spin inversion. For small $J$ the ground-state symmetry does not change. Then at $J=0.371$ the lowest-energy state acquires a finite $S=2$ and switches to the also singly-degenerate $A_g$ irreducible representation, with an $S^z=0$ component that is symmetric with respect to spin inversion. At $J=0.642$ it reverts to the low-$J$ ground state symmetry. For higher $J$ the ground state acquires again a finite $S$, as in the classical case. The ground-state $S$ value increases with $J$, and belongs to irreducible representations of spatial degeneracy higher than one. For $J \geq 1.685$ the ground state becomes again singly degenerate and belongs to the $A_g$ irreducible representation. It does not further change and up to the unfrustrated limit $J \to \infty$ has residual $S=4$.

\begin{figure}
\includegraphics[width=3.5in,height=2.5in]{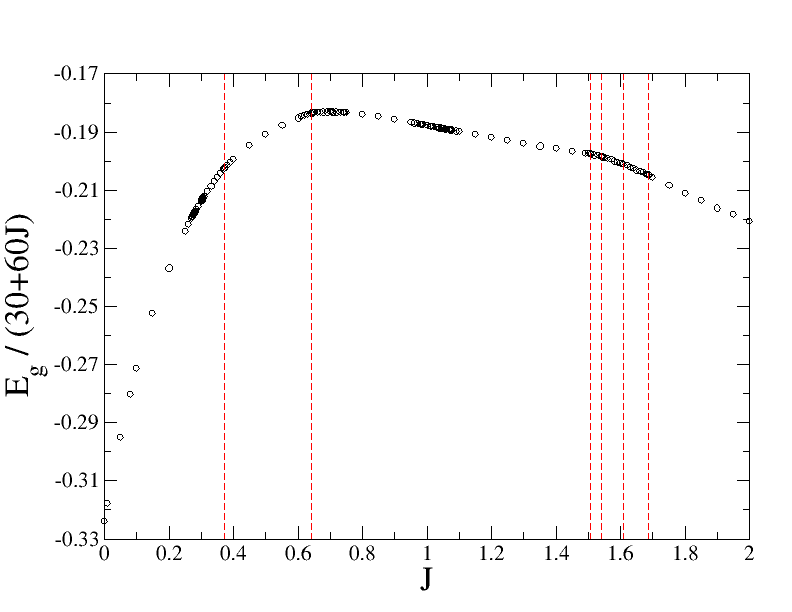}
\caption{(Color online) The (black) circles give the reduced energy of the zero-field ground state per bond $\frac{E_g}{30+60J}$ of the $s=\frac{1}{2}$ Hamiltonian (\ref{eqn:Hamiltonian}) plotted as a function of $J$. The limiting value for $J \to \infty$ is -0.33288 (Table \ref{table:lowenergyspectrumspinonehalf}). The (red) long-dashed vertical lines show the $J$ values where the total spin $S$ and the symmetry of the LEC change (Table \ref{table:spinonehalfzerofieldgroundstate}).
}
\label{fig:fig:pentakisdodecahedrongroundstateenergyspinonehalf}
\end{figure}

The strength of frustration as a function of $J$ is demonstrated by the plot of the reduced ground state energy per bond $\frac{E_g}{30+60J}$ (Fig. \ref{fig:fig:pentakisdodecahedrongroundstateenergyspinonehalf}). Frustration is strongest at $J=0.69881$, where the reduced ground state energy is maximized. This value of $J$ is somewhat smaller than the corresponding classical value.

\begin{figure}
\includegraphics[width=3.5in,height=2.5in]{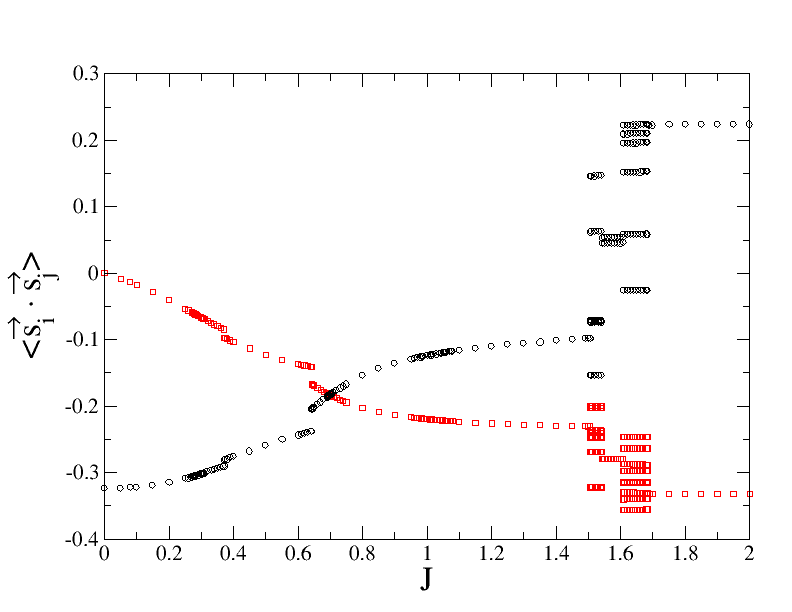}
\caption{(Color online) Nearest-neighbor correlations for the zero-field ground state of the $s=\frac{1}{2}$ Hamiltonian (\ref{eqn:Hamiltonian}) plotted as a function of $J$. The (black) circles give the intradodecahedron correlations, and the (red) squares the correlations between five-fold and six-fold coordinated spins. The $J$ values where the total spin $S$ and the symmetry of the LEC change are given in Table \ref{table:spinonehalfzerofieldgroundstate}.
}
\label{fig:spinonehalfgroundstatecorrelations}
\end{figure}

Fig. \ref{fig:spinonehalfgroundstatecorrelations} plots the nearest-neighbor correlations in the zero-field ground state. For $J=0$ only the intradodecahedral correlation is nonzero, and it is equal to $-0.32407$ \cite{NPK05}. As expected, with increasing $J$ the antiferromagnetic character of the correlation between six-fold and five-fold coordinated spins strengthens at the expense of the intradodecahedral one. At the point of maximum frustration $J=0.69881$ the two correlations become equal, with a value $-0.18307$. This is higher than the ground-state nearest-neighbor correlation for the icosahedron, which is equal to -0.20626 \cite{NPK05}, similarly to the classical case (Sec. \ref{subsec:groundstatezerofieldclassical}). In the unfrustrated limit $J \to \infty$ the nearest-neighbor correlations have values $0.23045$ and $-0.33288$, the latter being only slightly lower than the $J=0$ intradodecahedron frustrated correlation.

\subsection{Low-Energy Spectrum}
\label{sec:spinsonehalflowenergyspectrum}

The lowest-energy levels of Hamiltonian (\ref{eqn:Hamiltonian}) for $s=\frac{1}{2}$ are listed in Table \ref{table:lowenergyspectrumspinonehalf} for specific $J$ values. For $J=0$ only the dodecahedron spins interact and frustration results in six singlets inside the singlet-triplet gap \cite{NPK05}. Once $J$ becomes nonzero the icosahedron spins become coupled to the rest and each $J=0$ level gives rise to a band of $2^{12}$ eigenstates. These low-lying levels are triplets, as can be seen for $J=0.2$ and 0.3. Singlet excitations inside the singlet-triplet gap reemerge for $J=0.9$ and survive up to $J=1.2$, demonstrating that from the point of view of nonmagnetic excitations in the low-energy spectrum the effect of frustration is strongest when the two coupling constants are of roughly equal strength, with a maximum of 15 singlets in the singlet-triplet gap when $J=1.1$.



\subsection{Ground-State Magnetization Response in an External Field}
\label{subsec:groundstatemagnetizationresponseinanexternalfieldspinonehalf}

When the magnetic field $h$ in Hamiltonian ($\ref{eqn:Hamiltonian}$) becomes nonzero, the lowest levels in the various $S^z$ sectors determine the ground-state magnetization as $h$ is varied from zero to its saturation value. The energy of each level varies linearly with the field, in proportionality with its $S^z$ value. Typically, successive $S^z$ sectors include the ground state in increasing field, and for specific field values the ground state switches between neighboring $S^z$ sectors for which $\Delta S^z=1$. It may happen for frustrated molecules that the lowest energy in a specific $S^z$ sector never becomes the ground state in the field, in which case there is a quantum magnetization discontinuity with $\Delta S^z=2$ \cite{NPK05,NPK07,NPK16}. This can also happen for extended systems, not necessarily frustrated \cite{Schulenburg02,Nakano13,Nakano15}.

\begin{figure}
\includegraphics[width=3.5in,height=2.5in]{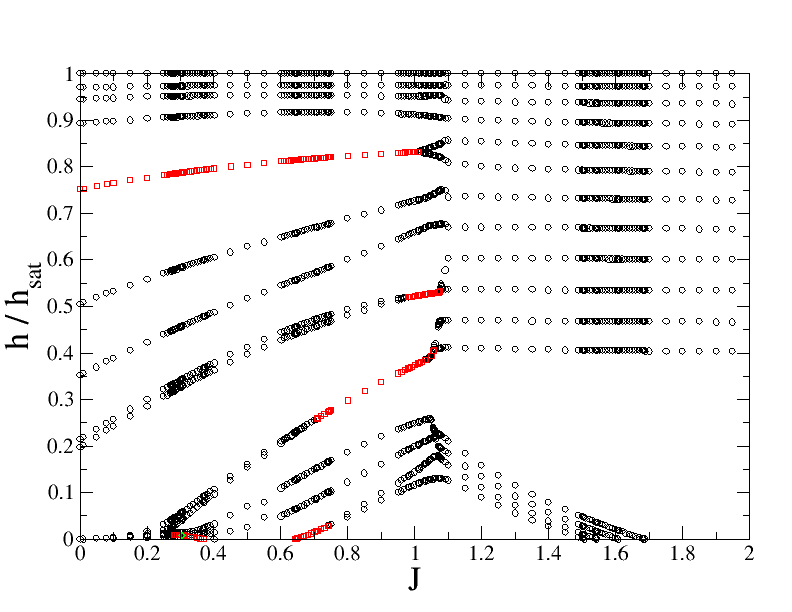}
\caption{(Color online) Magnetization discontinuities in the ground state of the $s=\frac{1}{2}$ Hamiltonian (\ref{eqn:Hamiltonian}) as a function of $J$ and the magnetic field $h$ over its saturation value $h_{sat}$. (Black) circles represent discontinuities with $\Delta S^z=1$, (red) squares discontinuities with $\Delta S^z=2$, and (green) diamonds discontinuities with $\Delta S^z=3$.
}
\label{fig:spinonehalfgroundstatemagnetization}
\end{figure}

\begin{figure}
\includegraphics[width=3.5in,height=2.5in]{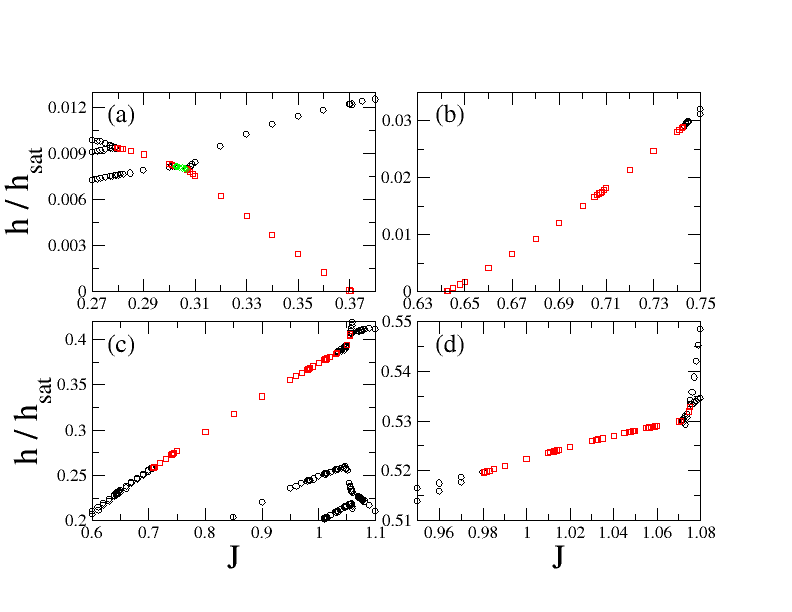}
\caption{(Color online) Parts of Fig. \ref{fig:spinonehalfgroundstatemagnetization} with discontinuities $\Delta S^z > 1$ shown in greater detail.
}
\label{fig:spinonehalfgroundstatemagnetizationfocus}
\end{figure}

Figures \ref{fig:spinonehalfgroundstatemagnetization} and \ref{fig:spinonehalfgroundstatemagnetizationfocus} show the discontinuities of the ground-state magnetization of Hamiltonian ($\ref{eqn:Hamiltonian}$) as a function of $J$ and $\frac{h}{h_{sat}}$. The $J$-range of the $\Delta S^z > 1$ discontinuities is listed in Table \ref{table:spinonehalfmagnetizationdiscontinuities}. The jump associated with the dodecahedron ($J=0$) survives up to $J \sim 1$, something similar to the classical case (Sec. \ref{subsec:groundstatemagnetizationresponseinanexternalfieldclassical}). The competition between the two exchange constants generates further discontinuities for low and higher field, with $\Delta S^z$ going up to 3. A $\Delta S^z = 3$ jump has never been found for fullerene-type molecules. The biggest number of $\Delta S^z > 1$ discontinuities for a specific $J$ equals 3 when $0.707 < J < 0.743$ and $0.980 \leq J \leq 1.012$. The total of three $\Delta S^z=2$ ground-state magnetization discontinuities for the same $J$ supersedes the total number of discontinuities found in the past for fullerene molecules. The dodecahedron has been found to have two discontinuities when $s=1$ \cite{NPK05}. Two of the magnetization discontinuities come one after the other as the field is increased for $0.980 \leq J \leq 1.032$ and also for $1.050 \leq J \leq 1.056$, something encountered only in the case of the two coupled dodecahedra \cite{NPK16}.

The $s=\frac{1}{2}$ and 1 dodecahedron discontinuities involve a change of ground-state symmetry from $A_g$ to $A_u$ with increasing field \cite{NPK05}.
The same mechanism governs the high-field discontinuity for other $I_h$-fullerene molecules \cite{NPK07}. Table \ref{table:spinonehalfmagnetizationdiscontinuities} shows that not only the reverse can happen in the pentakis dodecahedron, but also that degenerate irreducible representations can be the ground states on either side of a jump. Especially the discontinuity for $0.279 < J \leq 0.302$ involves only nondegenerate irreducible representations.

The transition from the highly frustrated toward the nonfrustrated $J \to \infty$ limit as a function of $J$ is also demonstrated in Fig. \ref{fig:spinonehalfgroundstatemagnetization}. For higher $J$ what remains in the ground-state magnetic response is a plateau that originates in the residual ground-state magnetization. Once this plateau is overcome by the field the magnetization jumps are roughly equidistant, showing that frustration is weak for higher $J$. On the other hand, for lower $J$ where frustration is stronger the increase in the field strength required to lead to the next magnetization jump varies significantly with $h$.

\subsection{Quantum Spins $s=1$}
\label{subsec:spinsone}

The calculation of the $s=1$ ground-state magnetization is limited to higher $S^z$ values due to computational requirements. The dodecahedron ($J=0$) has two $\Delta S^z=2$ discontinuities for $s=1$ \cite{NPK05}, but only the higher-field one can be traced as a function of increasing $J$. It is the discontinuity where the sector with five flipped spins never becomes the ground state in the field, just as in the $s=\frac{1}{2}$ case. It eventually disappears for $J=0.585$, a weaker value than the one for $s=\frac{1}{2}$.

\section{Conclusions}
\label{sec:conclusions}

The pentakis dodecahedron is the second-smallest fullerene dual molecule after the icosahedron. The presence of two inequivalent vertices necessitates the introduction of two different exchange constants when spins mounted on each vertex of the molecule interact according to the AHM. Investigation of the ground-state magnetic response as a function of the relative strength of the exchange constants has demonstrated its richness in classical magnetization and susceptibility discontinuities. The quantum response includes the sought-after magnetization discontinuities with $\Delta S^z=2$ and even $\Delta S^z=3$. When the exchange constants are comparable in strength a number of nonmagnetic excitations is found inside the singlet-triplet gap. The discontinuous magnetic response and the low-energy structure of the spectrum are consequences of the frustrated connectivity of the molecule.

The calculation of the magnetic properties of the pentakis dodecahedron is the first step to investigate dual fullerenes bigger than the icosahedron and search for symmetry patterns in their magnetic behavior. The presence of inequivalent types of bonds further complicates this search.

\begin{appendix}

\section{Classical Saturation $J$ in Zero Magnetic Field}
\label{appendix:saturationJzeromagneticfield}

When $J$ in Hamiltonian (\ref{eqn:Hamiltonian}) becomes strong enough, the five-fold coordinated spins become antiparallel to their nearest neighbors in the classical LEC, while the six-fold coordinated spins are parallel to each other (Fig. \ref{fig:pentakisdodecahedronzerofieldcorrelationscontributions}). Approaching the value of $J$ for which this saturation sets in from below (Figs \ref{fig:lowestenergyconfigurationdiscontinuities}(a) and \ref{fig:lowestenergyconfigurationanglesJ=1.08}), the LEC is CF8 (Fig. \ref{fig:lowestenergyconfigurations}), which has two of the five-fold coordinated spins antialigned with the total magnetization, which is taken along the $z$ axis. The rest of the spins are divided in three groups of ten, the two corresponding to six-fold and the other to five-fold coordinated spins, where in each group the polar angle is the same. For these thirty spins there are ten unique equidistantly spaced azimuthal angles, with each corresponding to three spins. Each azimuthal angle includes a spin from each of the three families of polar angle spins. This configuration has both the dodecahedron and the rest of the spins resembling the high-field ground states of the dodecahedron and the icosahedron respectively \cite{NPK07,Schroeder05,NPK15}. Its energy functional is

\begin{eqnarray}
\frac{E}{10}-2 & = & J C \textrm{cos}\theta_1 - \frac{\sqrt{5}+5}{4} \textrm{sin}^2\theta_1 - \frac{\sqrt{5}+3}{4} \textrm{sin}^2\theta_2 + \nonumber \\ & & \textrm{cos}(\theta_1+\theta_2) + J \textrm{cos}(\theta_2+\theta_3) + \nonumber \\ & & J ( \frac{\sqrt{5}-1}{2} \textrm{sin}\theta_2 \textrm{sin}\theta_3 + 2 \textrm{cos}\theta_2 \textrm{cos}\theta_3 ) + \nonumber \\ & & J ( -\frac{\sqrt{5}-1}{2} \textrm{sin}\theta_1 \textrm{sin}\theta_3 + 2 \textrm{cos}\theta_1 \textrm{cos}\theta_3 )
\label{eqn:appendixfirst}
\end{eqnarray}

with $C=-1$. Approaching the $J$ value for which the spins become colinear the polar angles of the dodecahedron spins $\theta_1,\theta_2 \to 0$, while the polar angle of the rest of the spins $\theta_3 \to \pi$ (Fig. \ref{fig:pentakisdodecahedronzerofieldmagnetizationcontributions}). Making the variable change $\theta_3 \to \pi - \theta_3$ all polar angles are small at this limit and a small angles expansion yields

\begin{eqnarray}
\frac{E}{10}-3+6J & \approx & \frac{1}{2} ( 3J-\frac{\sqrt{5}+7}{2} ) \theta^2_1 + \nonumber \\ & & \frac{1}{2} ( 3J-\frac{\sqrt{5}+5}{2} ) \theta^2_2 + \frac{5}{2} J \theta^2_3 - \theta_1 \theta_2 + \nonumber \\ & & \frac{\sqrt{5}-3}{2} J \theta_2 \theta_3 - \frac{\sqrt{5}-1}{2} J \theta_1 \theta_3
\end{eqnarray}

Taking the derivatives with respect to the three polar angles gives

\begin{eqnarray}
\frac{1}{10} \frac{\partial E}{\partial \theta_1} & \approx & ( 3J-\frac{\sqrt{5}+7}{2} ) \theta_1 - \theta_2 - \frac{\sqrt{5}-1}{2}J \theta_3 \nonumber \\
\frac{1}{10} \frac{\partial E}{\partial \theta_2} & \approx & - \theta_1 + ( 3J - \frac{\sqrt{5}+5}{2} ) \theta_2 + \frac{\sqrt{5}-3}{2} J \theta_3 \nonumber \\
\frac{1}{10} \frac{\partial E}{\partial \theta_3} & \approx & - \frac{\sqrt{5}-1}{2} J \theta_1 + \frac{\sqrt{5}-3}{2} J \theta_2 + 5J \theta_3
\end{eqnarray}

To find the energy minimum the three derivatives must be set equal to zero, and for a solution to exist the determinant of the coefficients of the three angles must equal zero. This is equivalent to $J=0$ and the equation
\begin{eqnarray}
2(5+\sqrt{5}) J^2 - (25+7\sqrt{5})J + 5(3+\sqrt{5}) = 0
\end{eqnarray}
One of the two roots of this equation is the value that brings saturation in zero magnetic field, $J=\frac{5+\sqrt{5}}{4}$.

\section{Saturation Magnetic Field}
\label{appendix:saturationmagneticfield}

The saturation field can be calculated in two ways, for quantum-mechanical and for classical spins.

\subsection{Quantum-Mechanical Spins}

For arbitrary $s$, the lowest-energy state in the sector a single spin-flip away from saturation belongs to the $T_{2u}$ irreducible representation for lower $J$, and to the $A_g$ representation for higher $J$. If $|i>$ is a state where all spins are up except for the one at site $i$, with sites 1 to 12 belonging to the icosahedron and 13 to 32 to the dodecahedron, the two basis states for one of the three Hamiltonian subblocks belonging to the $T_{2u}$ representation are:
\begin{eqnarray}
|\Psi_1^{T_{2u}}> & = & \frac{5-\sqrt{5}}{20} ( |13> + |14> - |31> - |32> ) - \nonumber \\ & & \frac{\sqrt{5}}{10} ( |15> + |16> + |17> + |18> - |27> - \nonumber \\ & & |28> - |29> - |30> ) + \frac{5+\sqrt{5}}{20} ( |19> + \nonumber \\ & & |20> - |25> - |26> ) \nonumber \\
|\Psi_2^{T_{2u}}> & = & \frac{\sqrt{5-\sqrt{5}}}{2\sqrt{10}} ( |1> + |2> - |11> - |12> ) - \nonumber \\ & & \frac{\sqrt{5+\sqrt{5}}}{2\sqrt{10}} ( |3> + |4> - |9> - |10> )
\end{eqnarray}
The subblock of the Hamiltonian matrix corresponding to these two states is, with the energy of the state with all spins up $E_{sat}=30(1+2J)s^2$

\begin{displaymath}
\left(
\begin{array}{cc}
E_{sat}-3(1+J)s-\sqrt{5}s & \sqrt{5-2\sqrt{5}} J s \\
\sqrt{5-2\sqrt{5}} J s & E_{sat}-5Js \\
\end{array} \right)
\end{displaymath}

The saturation field for lower $J$ is the difference between $E_{sat}$ and the lowest eigenvalue of this Hamiltonian subblock, and it is equal to $\frac{1}{2}[3+\sqrt{5}+8J+\sqrt{(3+\sqrt{5})^2-4(3+\sqrt{5})J+8(3-\sqrt{5})J^2}]s$.

The two basis states for the single Hamiltonian subblock belonging to the $A_g$ representation are:
\begin{eqnarray}
|\Psi_1^{A_g}> & = & \frac{1}{2\sqrt{5}} \sum_{i=13}^{32} |i> \nonumber \\
|\Psi_2^{A_g}> & = & \frac{1}{2\sqrt{3}} \sum_{i=1}^{12} |i>
\end{eqnarray}
The subblock of the Hamiltonian matrix corresponding to these two states is

\begin{displaymath}
\left(
\begin{array}{cc}
E_{sat}-3Js & \sqrt{15} J s \\
\sqrt{15} J s & E_{sat}-5Js \\
\end{array} \right)
\end{displaymath}

The saturation field for higher $J$ is the difference between $E_{sat}$ and the lowest eigenvalue of this Hamiltonian subblock, and it is equal to $8Js$.

The transition between the two saturation fields occurs when they are equal, which is when $J$ has the value $\frac{3}{20}(5+\sqrt{5})$.

\subsection{Classical Spins}

For lower $J$ the classical LEC after the last magnetization discontinuity and before saturation (Figs \ref{fig:lowestenergyconfigurationdiscontinuities}(a), \ref{fig:lowestenergyconfigurationdiscontinuities}(c), \ref{fig:lowestenergyconfigurationdiscontinuitiesfocus}(l), \ref{fig:lowestenergyconfigurationanglesJ=0.3}, and \ref{fig:lowestenergyconfigurationanglesJ=1}) is CF6 (Fig. \ref{fig:lowestenergyconfigurations}).
This is the same with the configuration around the total magnetization in App. \ref{appendix:saturationJzeromagneticfield}, with the only difference that the two five-fold coordinated spins are parallel (to the magnetic field) and not antiparallel (to the total magnetization). The energy functional is (\ref{eqn:appendixfirst}) with $C=1$ plus the magnetic energy
\begin{eqnarray}
- 2h ( 1 + 5 \textrm{cos}\theta_1 + 5 \textrm{cos}\theta_2 + 5 \textrm{cos}\theta_3 )
\end{eqnarray}
Performing a small angles approximation and taking the derivatives with respect to the three polar angles gives
\begin{eqnarray}
\frac{1}{10} \frac{\partial E}{\partial \theta_1} & \approx & - ( 3J+\frac{\sqrt{5}+7}{2} - h ) \theta_1 - \theta_2 - \frac{\sqrt{5}-1}{2} J \theta_3 \nonumber \\
\frac{1}{10} \frac{\partial E}{\partial \theta_2} & \approx & - \theta_1 - ( 3J+\frac{5+\sqrt{5}}{2} - h ) \theta_2 + \frac{\sqrt{5}-3}{2} J \theta_3 \nonumber \\
\frac{1}{10} \frac{\partial E}{\partial \theta_3} & \approx & - \frac{\sqrt{5}-1}{2} J \theta_1 + \frac{\sqrt{5}-3}{2} J \theta_2 - ( 5J - h ) \theta_3
\end{eqnarray}
The energy minimum satisfies the equation
\begin{eqnarray}
& & h^3 - (6+\sqrt{5}+11J) h^2 + [3(3+\sqrt{5})+8(6+\sqrt{5})J+ \nonumber \\ & & 2(17+\sqrt{5})J^2]h -15(3+\sqrt{5})J-3(25+7\sqrt{5})J^2- \nonumber \\ & & 6(5+\sqrt{5})J^3=0
\end{eqnarray}
$h=3(1+J)$ is one of the roots, with the other two solutions of the equation
\begin{eqnarray}
h^2 - (3+\sqrt{5}+8J) h + 5(3+\sqrt{5})J + 2(5+\sqrt{5})J^2 = 0
\end{eqnarray}

One of the two roots is the saturation field $\frac{1}{2}[3+\sqrt{5}+8J+\sqrt{(3+\sqrt{5})^2-4(3+\sqrt{5})J+8(3-\sqrt{5})J^2}]$.

For higher $J$, for fields between susceptibility discontinuity $25'$ and saturation (Fig. \ref{fig:lowestenergyconfigurationdiscontinuities}(a)) there are two groups of spins in the classical LEC. The first includes the twenty six-fold coordinated spins, and the second the twelve five-fold coordinated spins. The spins in each group share the same polar and azimuthal angle, and the azimuthal angles of the two groups differ by $\pi$. The energy functional is

\begin{eqnarray}
\frac{E}{4} - \frac{15}{2} = 15 J \textrm{cos}(\theta_1+\theta_2) - h (5 \textrm{cos}\theta_1 + 3\textrm{cos}\theta_2)
\label{eqn:appendixsecond}
\end{eqnarray}
Close to saturation $\theta_1 \to 0$ and $\theta_2 \to 0$, and after a small angles approximation the energy functional becomes
\begin{eqnarray}
\frac{E}{4} - \frac{15}{2} - 15 J + 8 h & \approx & - \frac{15}{2}J (\theta_1+\theta_2)^2 + \frac{5}{2}h\theta_1^2+ \nonumber \\ & & \frac{3}{2}h\theta_2^2
\end{eqnarray}

Taking the derivatives with respect to the two unique polar angles yields
\begin{eqnarray}
\frac{1}{4} \frac{\partial E}{\partial \theta_1} & \approx & -5(3J-h)\theta_1-15J\theta_2 \nonumber \\
\frac{1}{4} \frac{\partial E}{\partial \theta_2} & \approx & -15J\theta_1-3(5J-h)\theta_2
\end{eqnarray}

The energy minimum is given by the equation
\begin{eqnarray}
h(h-8J)=0
\end{eqnarray}
The saturation field is $8J$.


\section{Classical Lower-Field Susceptibility Discontinuity for Higher $J$}
\label{appendix:classicallowerfieldmagnetizationdiscontinuityhigherJ}

For higher $J$ classical susceptibility discontinuity $24'$  (Fig. \ref{fig:lowestenergyconfigurationdiscontinuities}(a)) leads to a colinear configuration, with the six-fold coordinated spins parallel with the field and the five-fold antiparallel to it.
For fields below the discontinuity the LEC around the magnetic field is CF8 (Fig. \ref{fig:lowestenergyconfigurations}), which is the same with the configuration around the total magnetization in App. \ref{appendix:saturationJzeromagneticfield}, and the energy functional is (\ref{eqn:appendixfirst}) plus the magnetic energy
\begin{eqnarray}
- 2h ( -1 + 5 \textrm{cos}\theta_1 + 5 \textrm{cos}\theta_2 + 5 \textrm{cos}\theta_3 )
\end{eqnarray}
Making the change of variable $\theta_3 \to \pi - \theta_3$, performing a small angles approximation as the discontinuity is approached and taking the derivatives with respect to the three polar angles gives
\begin{eqnarray}
\frac{1}{10} \frac{\partial E}{\partial \theta_1} & \approx & ( 3J - \frac{\sqrt{5}+7}{2} - h ) \theta_1 - \theta_2 - \frac{\sqrt{5}-1}{2} J \theta_3 \nonumber \\
\frac{1}{10} \frac{\partial E}{\partial \theta_2} & \approx & - \theta_1 + ( 3J - \frac{\sqrt{5}+5}{2} - h ) \theta_2 + \frac{\sqrt{5}-3}{2} J \theta_3 \nonumber \\
\frac{1}{10} \frac{\partial E}{\partial \theta_3} & \approx & - \frac{\sqrt{5}-1}{2} J \theta_1 + \frac{\sqrt{5}-3}{2} J \theta_2 + ( 5J - h ) \theta_3
\end{eqnarray}
The energy minimum is given by the equation
\begin{eqnarray}
& & h^3 - (6+\sqrt{5}-J) h^2 + [3(3+\sqrt{5})+2(6+\sqrt{5})J- \nonumber \\ & & 2(8+\sqrt{5})J^2]h -15(3+\sqrt{5})J+3(25+7\sqrt{5})J^2- \nonumber \\ & & 6(5+\sqrt{5})J^3=0
\end{eqnarray}
$h=3(1-J)$ is one of the roots, with the other two solutions of the equation
\begin{eqnarray}
h^2 - (3+\sqrt{5}+2J) h + 5(3+\sqrt{5})J - 2(5+\sqrt{5})J^2 = 0
\end{eqnarray}
One of the two roots is the discontinuity field $h_{d1}=\frac{1}{2}[3+\sqrt{5}+2J-\sqrt{(3+\sqrt{5})^2-16(3+\sqrt{5})J+4(11+2\sqrt{5})J^2}]$. It becomes equal to zero when $J=\frac{5+\sqrt{5}}{4}$, in agreement with App. \ref{appendix:saturationJzeromagneticfield}, and for higher $J$ the discontinuity disappears.

\section{Classical Higher-Field Susceptibility Discontinuity for Higher $J$}
\label{appendix:classicalhigherfieldmagnetizationdiscontinuityhigherJ}

\begin{figure}
\includegraphics[width=3.5in,height=2.5in]{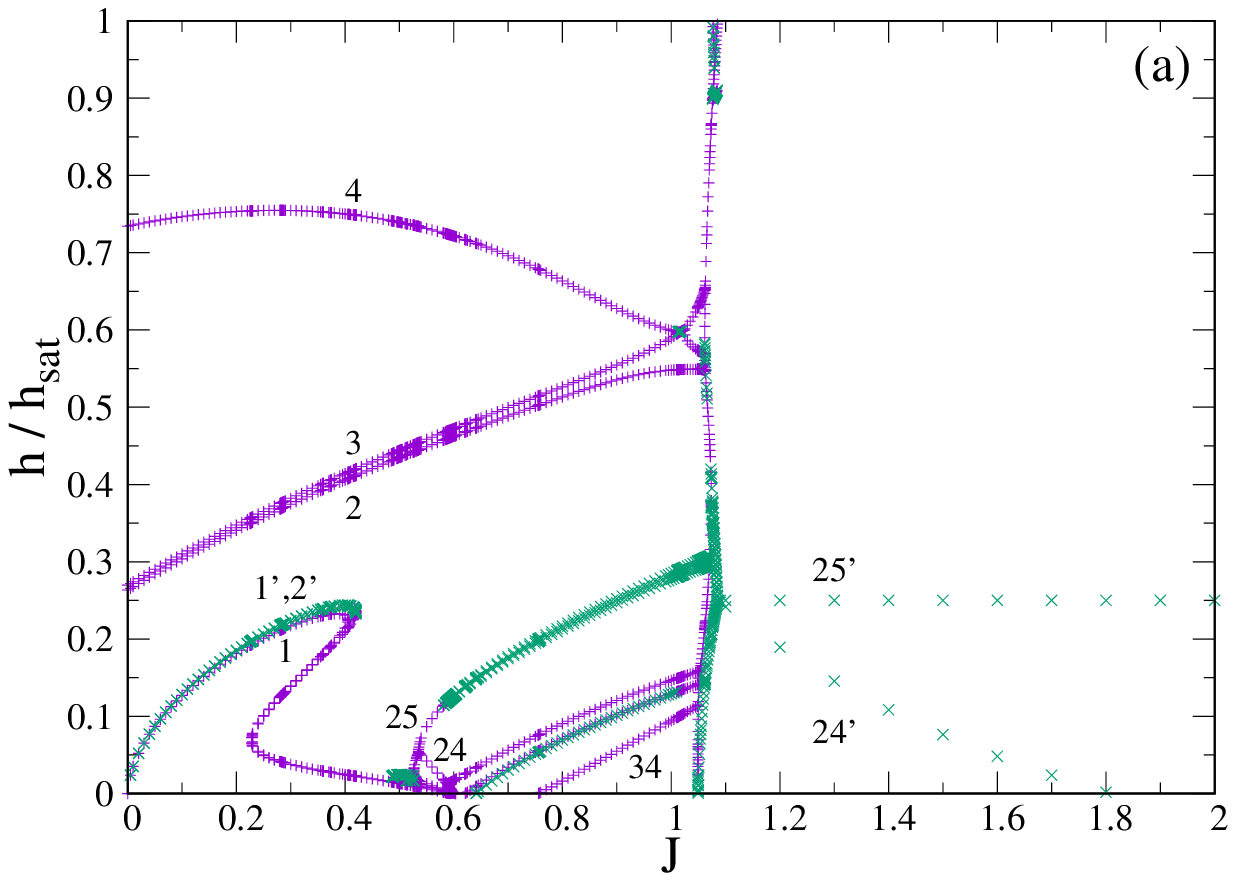}
\includegraphics[width=3.5in,height=2.5in]{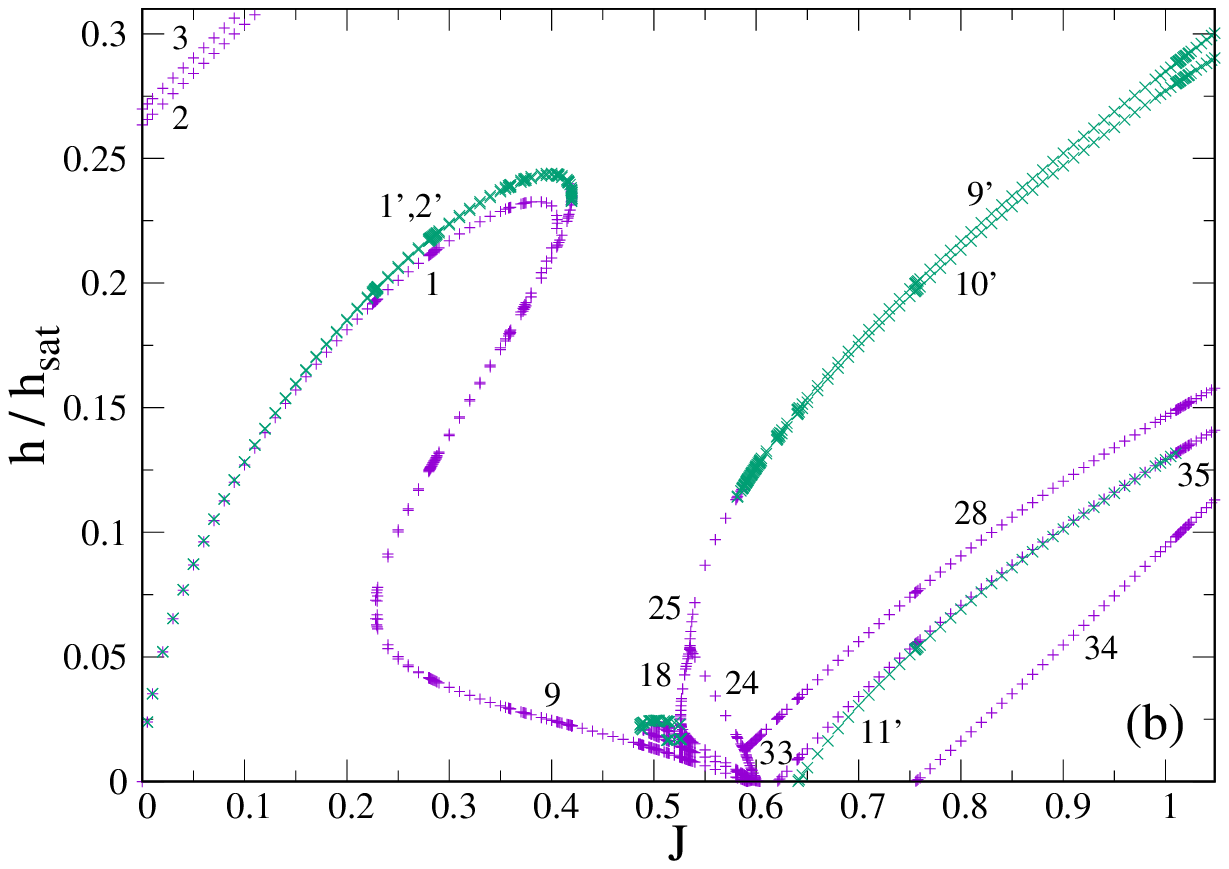}
\includegraphics[width=3.5in,height=2.5in]{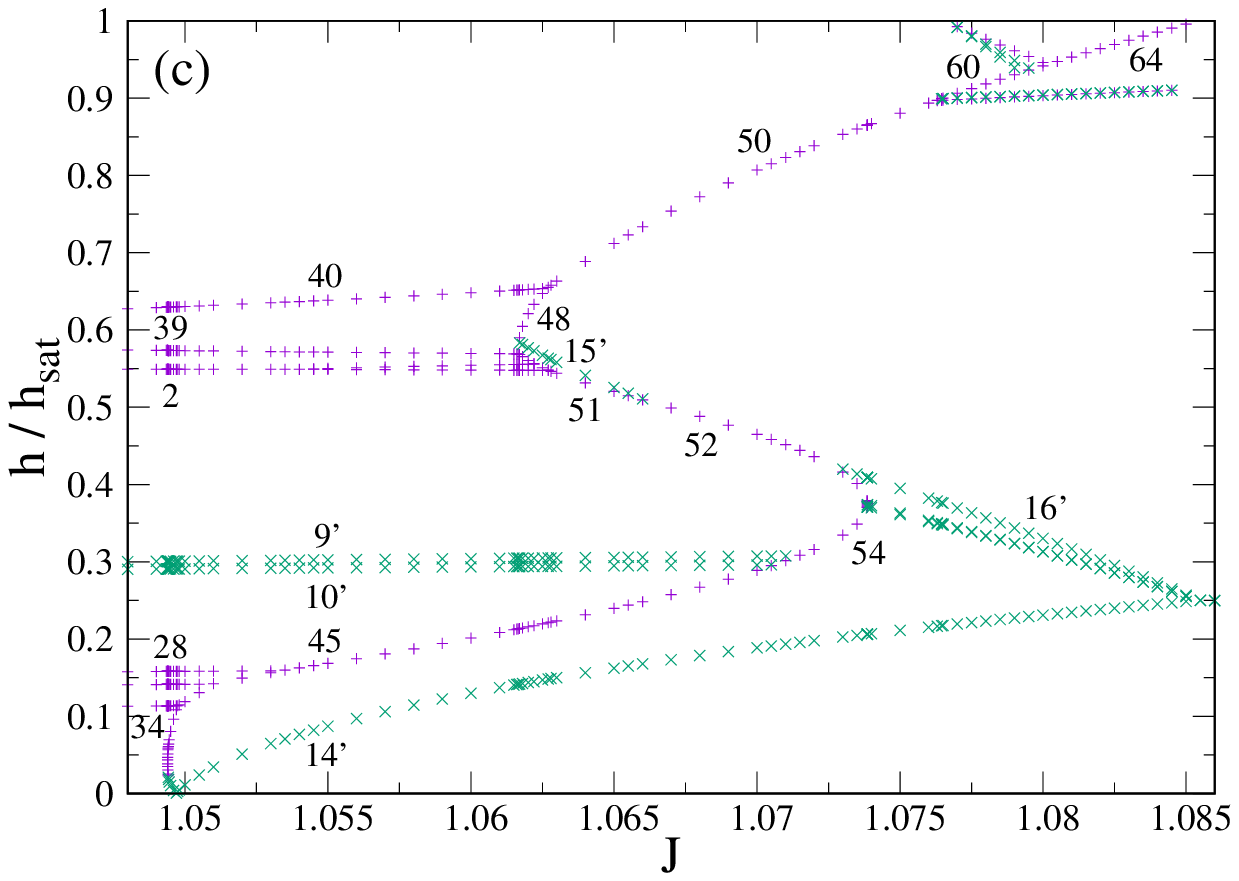}
\caption{(Color online) Location of magnetization and susceptibility discontinuties in the classical LEC of Hamiltonian (\ref{eqn:Hamiltonian}) as a function of $J$ and the magnetic field over its saturation value $\frac{h}{h_{sat}}$. (Purple) +'s are the magnetization, and (green) x's the susceptibility discontinuities. The numbers correspond to the discontinuities in Table \ref{table:classicalmagnsuscdisc}.
}
\label{fig:lowestenergyconfigurationdiscontinuities}
\end{figure}
\begin{figure*}
\centerline{
\includegraphics[width=0.37\textwidth,height=2.1in]{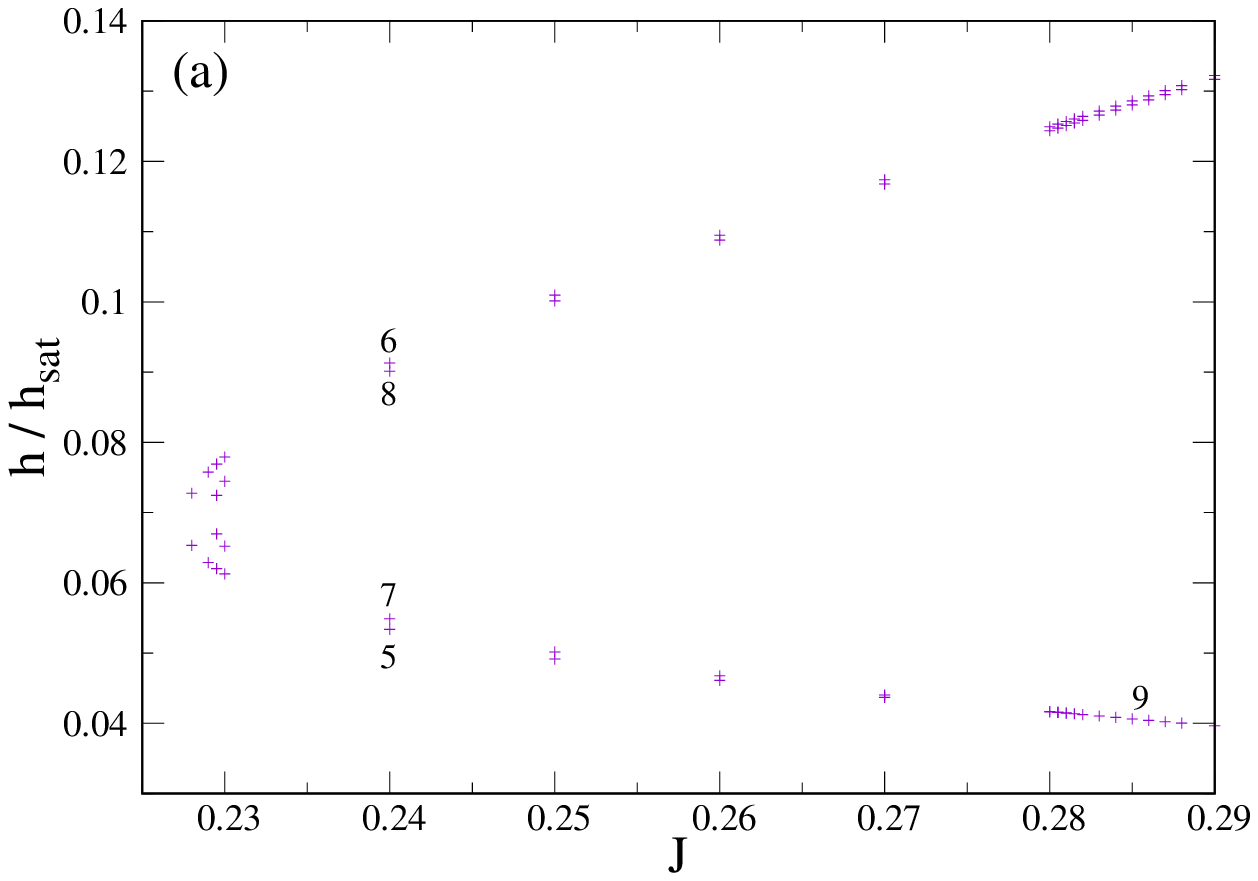}
\includegraphics[width=0.37\textwidth,height=2.1in]{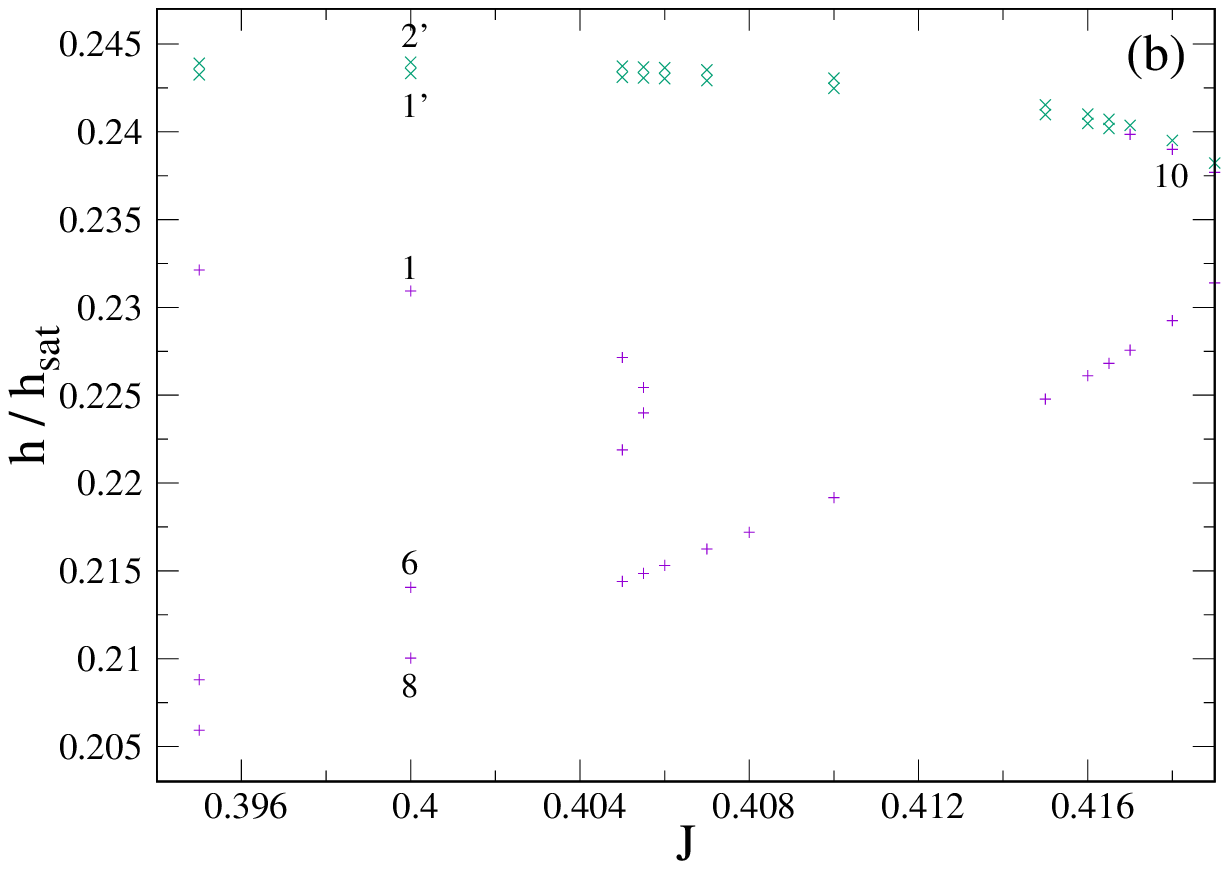}
\includegraphics[width=0.37\textwidth,height=2.1in]{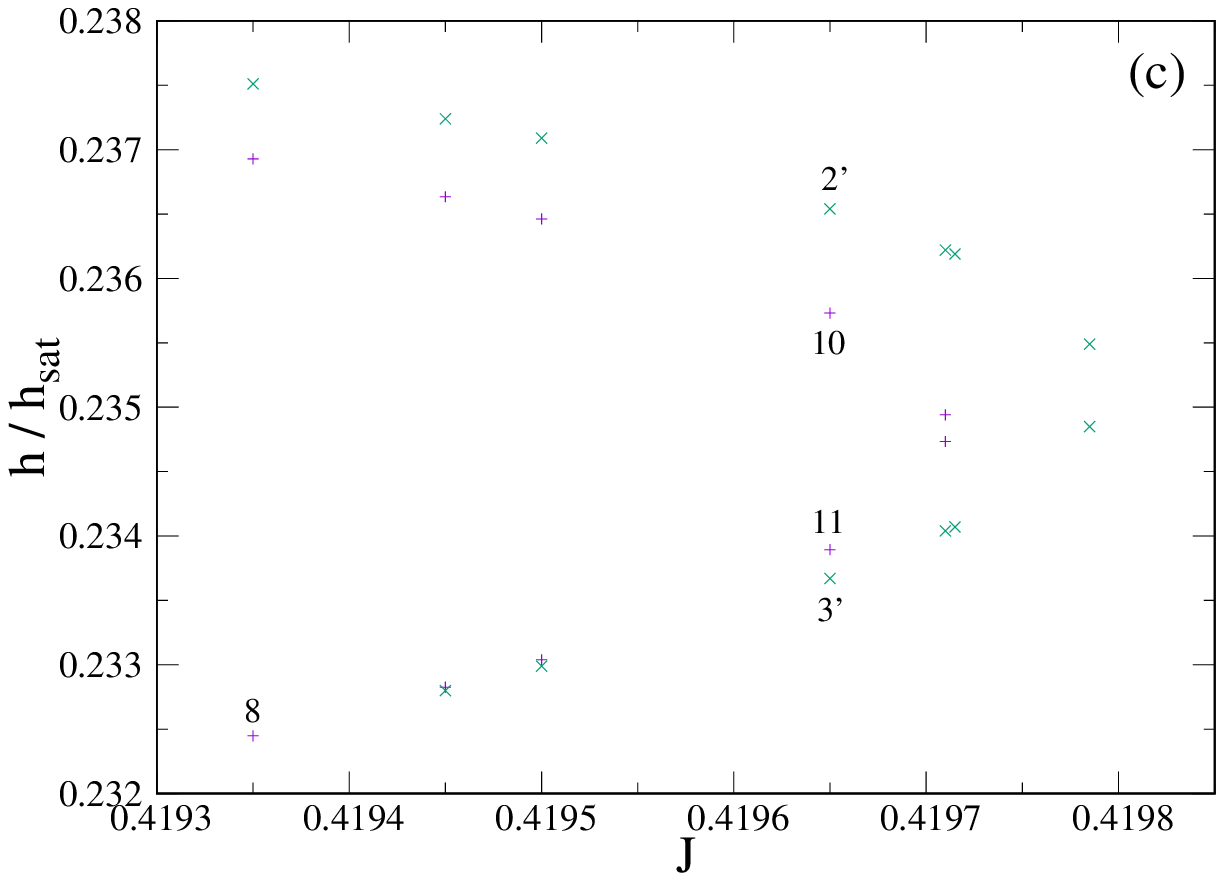}
}
\centerline{
\includegraphics[width=0.37\textwidth,height=2.1in]{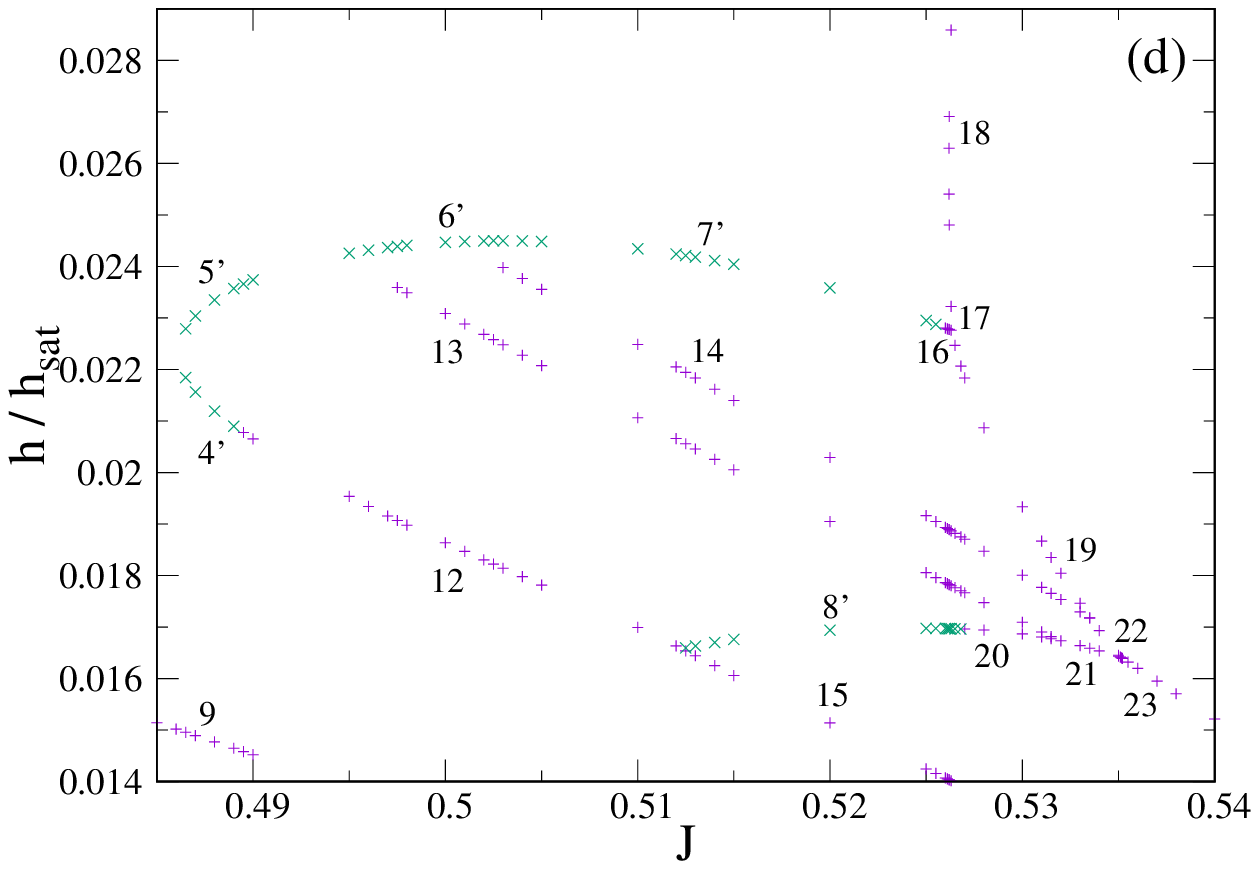}
\includegraphics[width=0.37\textwidth,height=2.1in]{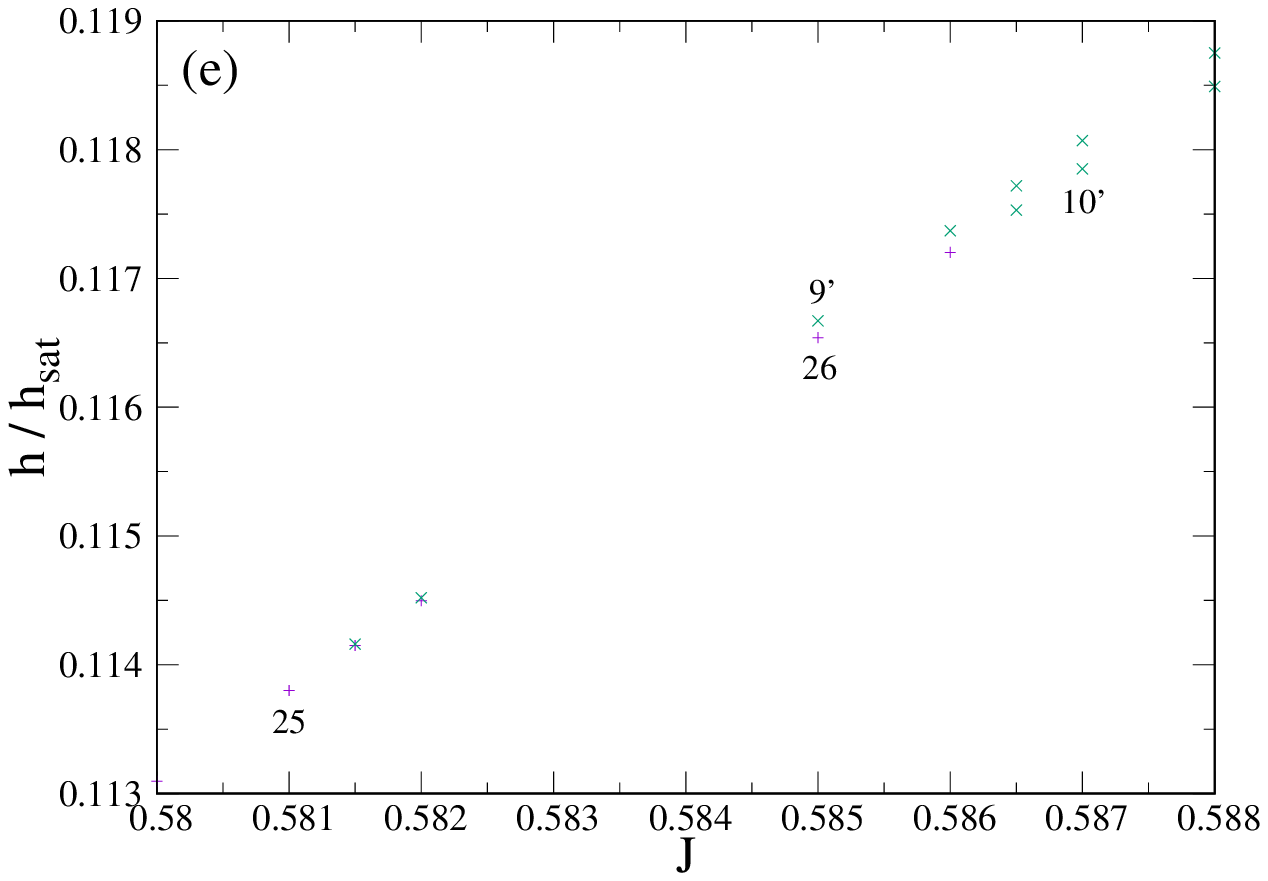}
\includegraphics[width=0.37\textwidth,height=2.1in]{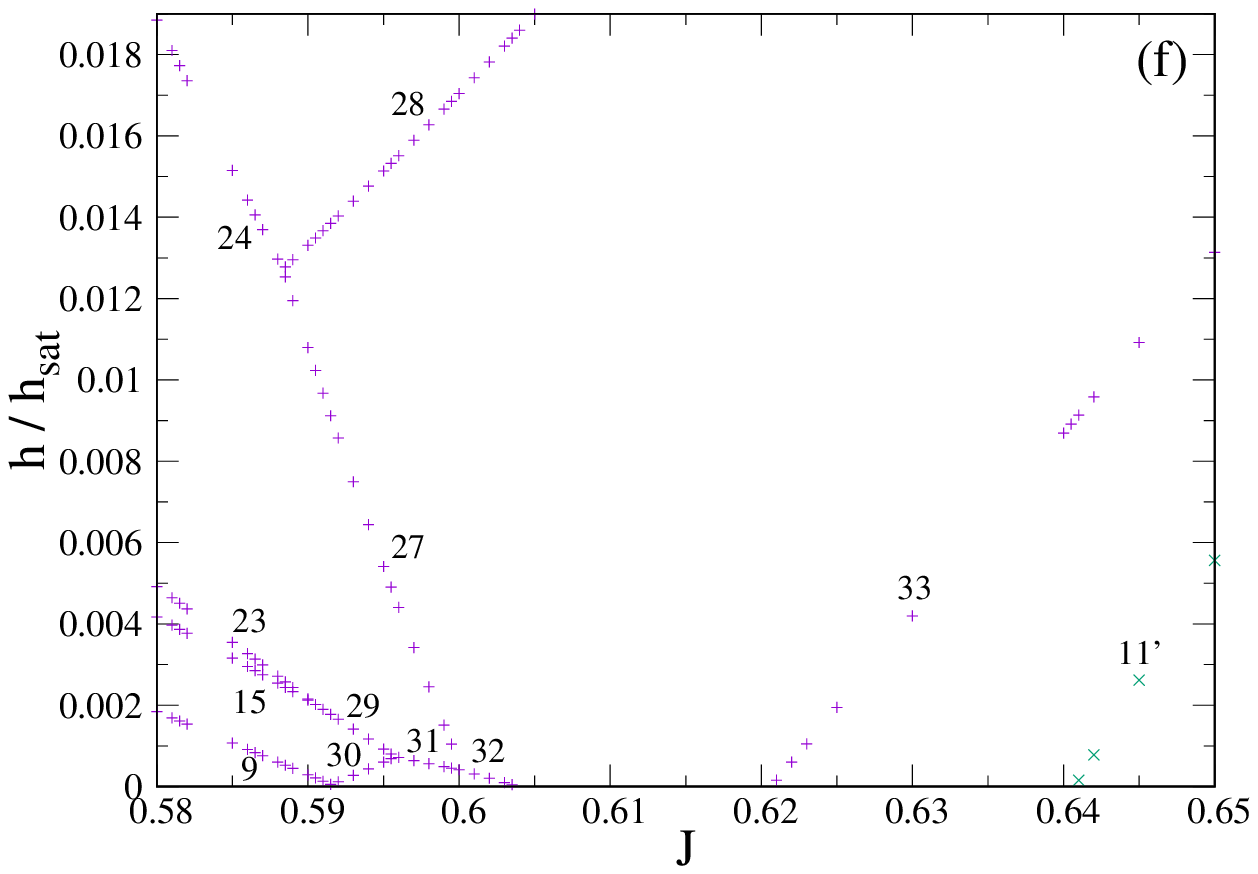}
}
\centerline{
\includegraphics[width=0.37\textwidth,height=2.1in]{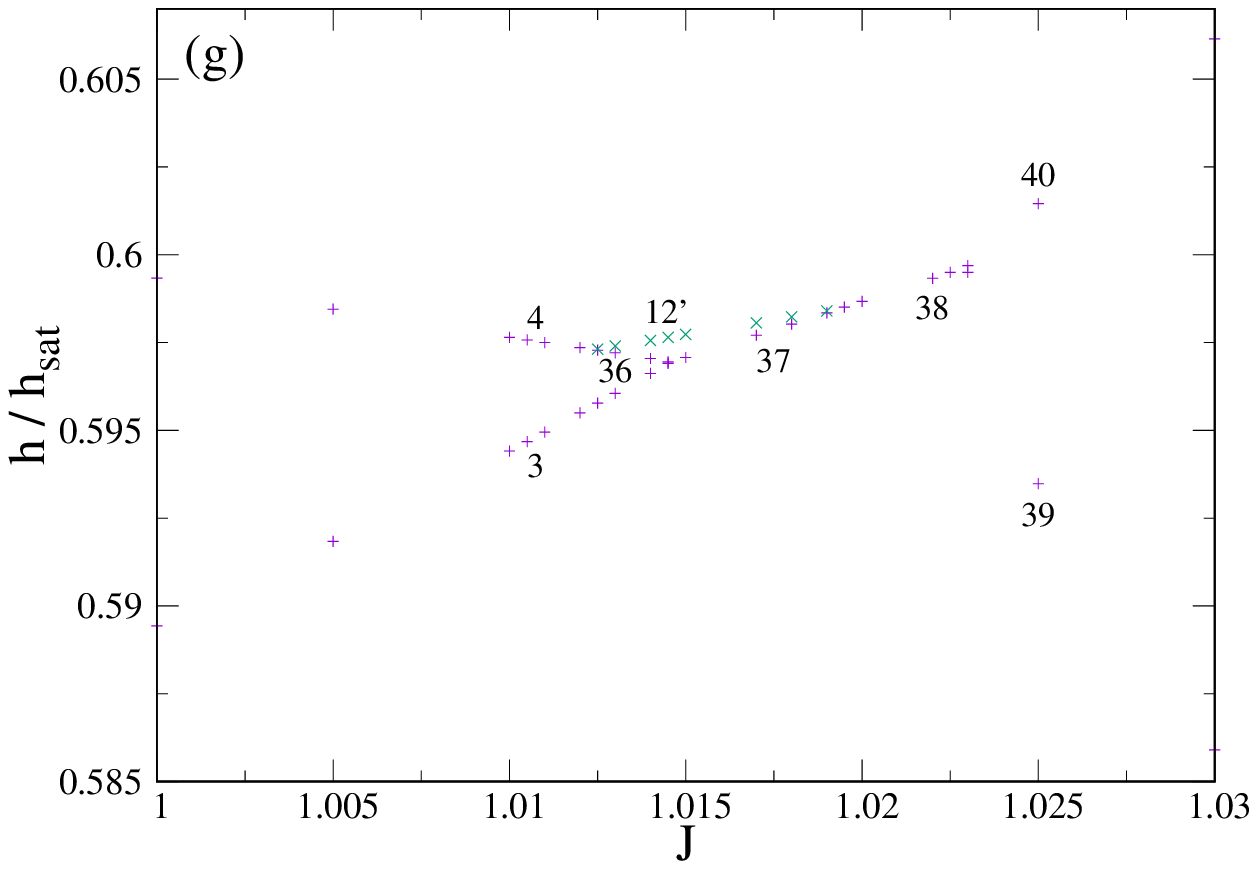}
\includegraphics[width=0.37\textwidth,height=2.1in]{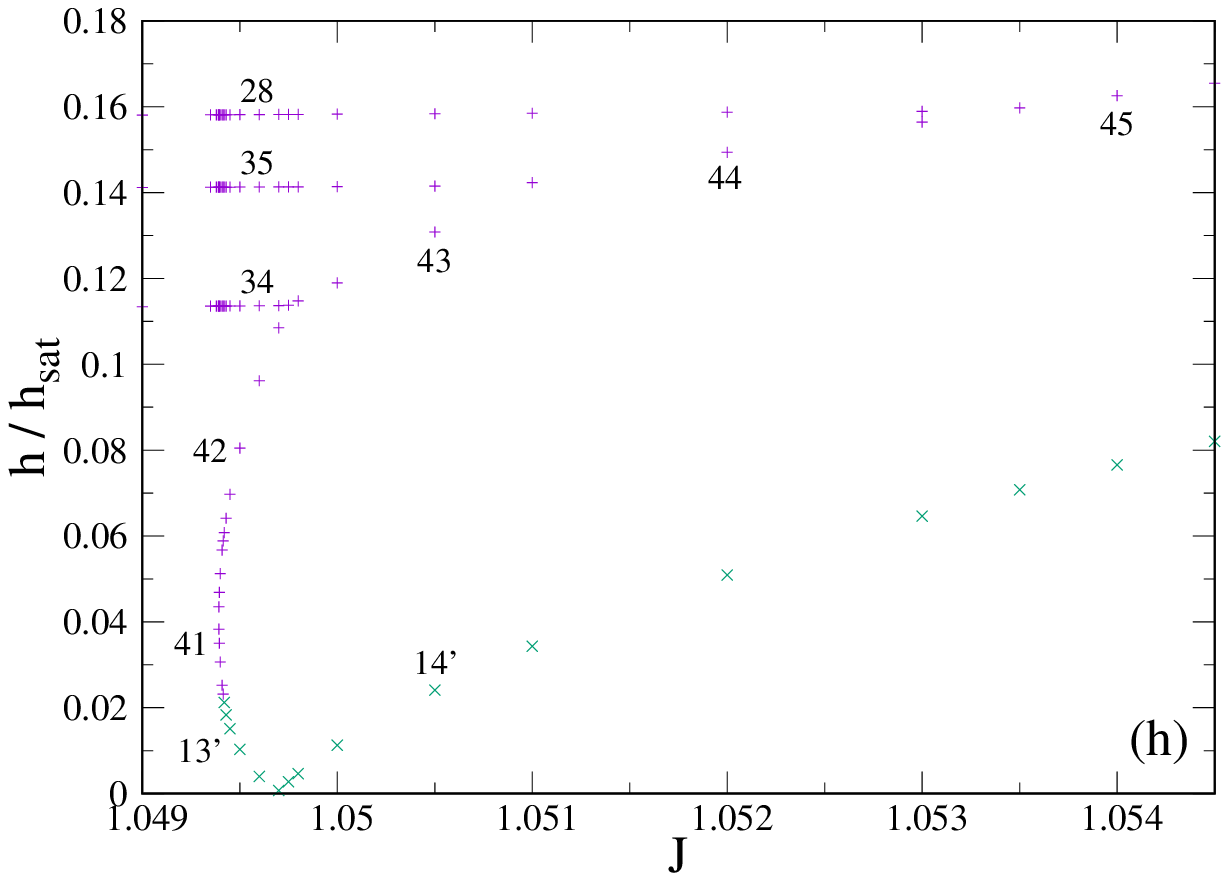}
\includegraphics[width=0.37\textwidth,height=2.1in]{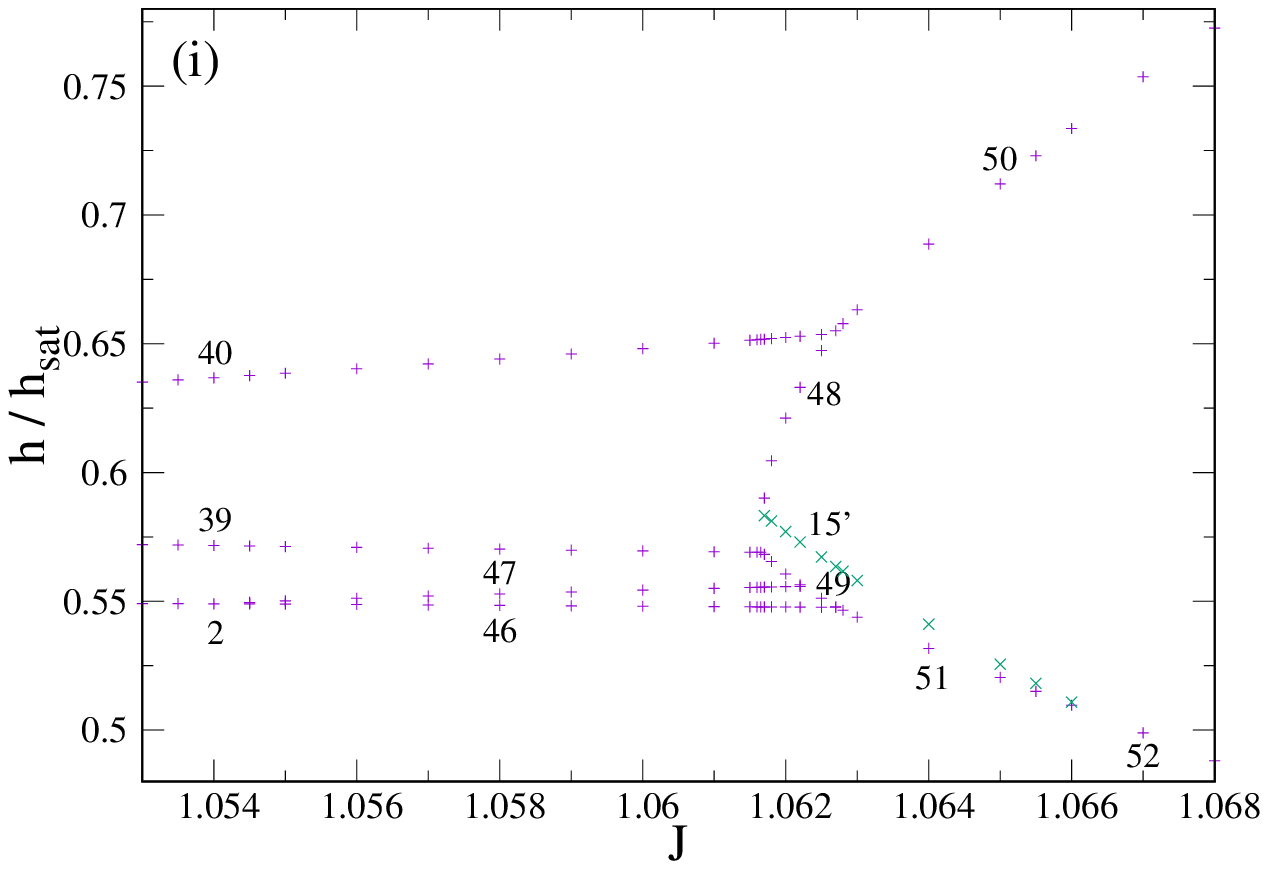}
}
\centerline{
\includegraphics[width=0.37\textwidth,height=2.1in]{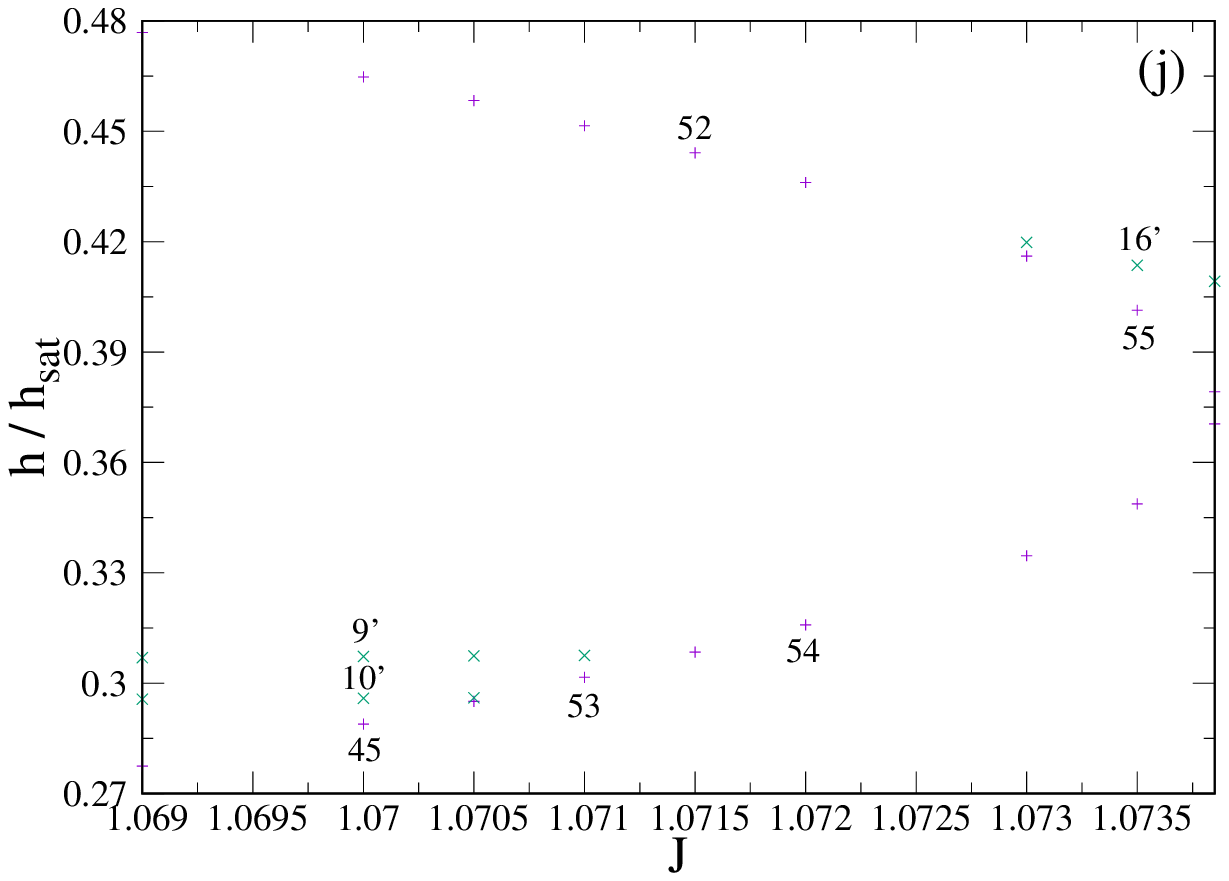}
\includegraphics[width=0.37\textwidth,height=2.1in]{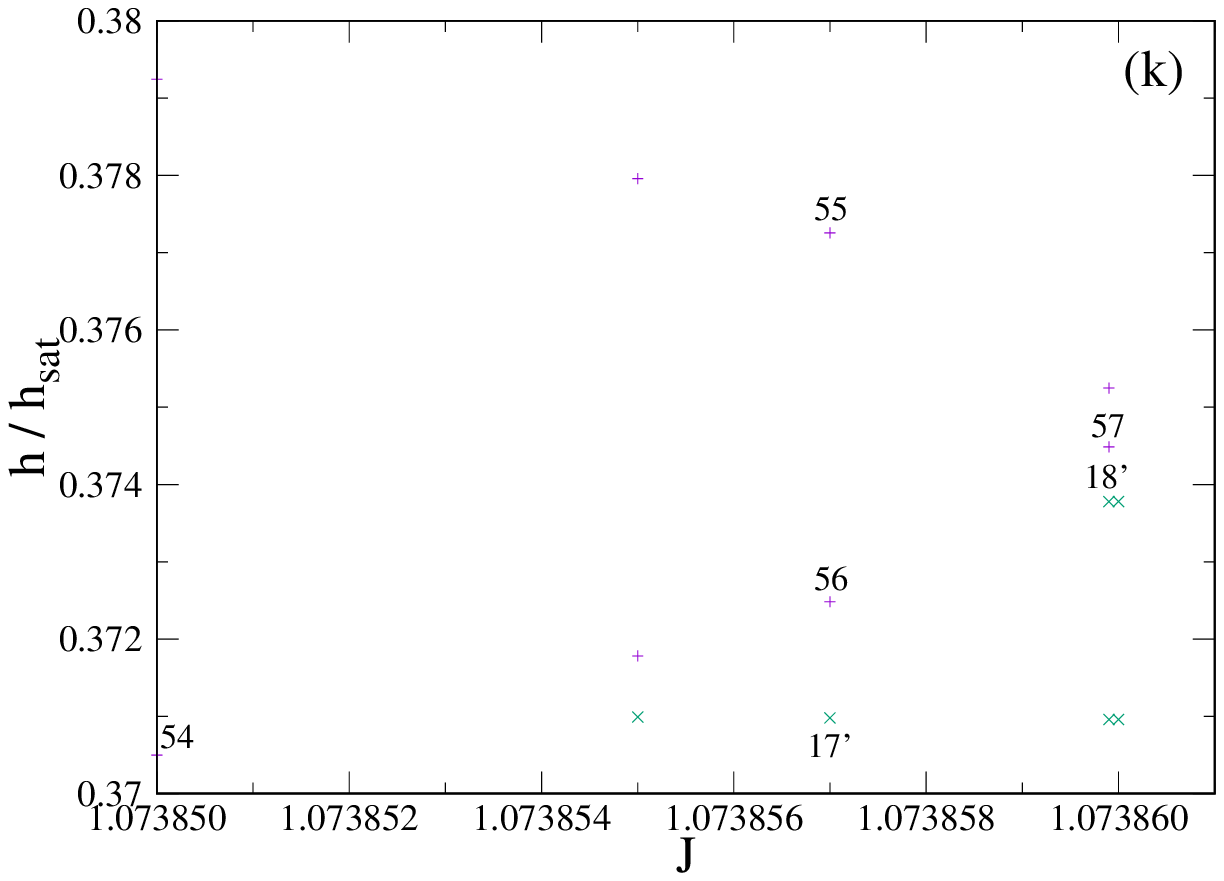}
\includegraphics[width=0.37\textwidth,height=2.1in]{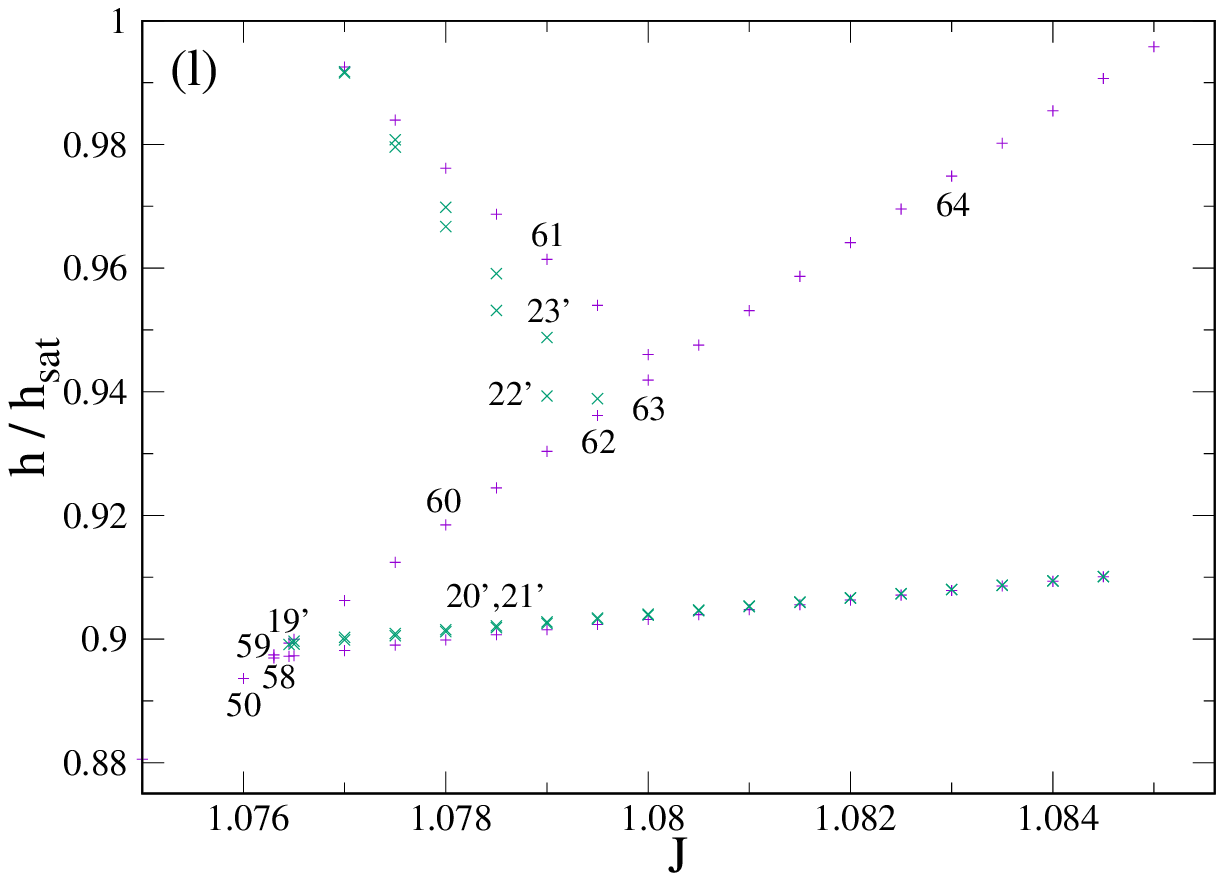}
}
\caption{(Color online) Parts of Fig. \ref{fig:lowestenergyconfigurationdiscontinuities} with discontinuities appearing and disappearing shown in greater detail.
}
\label{fig:lowestenergyconfigurationdiscontinuitiesfocus}
\end{figure*}

For higher $J$ and fields below the classical susceptibility discontinuity $25'$ (Fig. \ref{fig:lowestenergyconfigurationdiscontinuities}(a)) the spins are colinear, with the six-fold coordinated spins parallel with the field and the five-fold antiparallel to it. Between the discontinuity and saturation the LEC functional is given by (\ref{eqn:appendixsecond}). Making the variable change $\theta_2 \to \pi - \theta_2$, a small angles expansion as the discontinuity is approached from higher fields gives

\begin{eqnarray}
\frac{E}{4} - \frac{15}{2} + 15J + 2h & \approx & \frac{15J}{2} (\theta_1-\theta_2)^2 + \frac{h}{2} ( 5 \theta_1^2 - 3 \theta_2^2 )
\end{eqnarray}

Taking the derivatives with respect to the two polar angles yields

\begin{eqnarray}
\frac{1}{4} \frac{\partial E}{\partial \theta_1} & \approx &5 (3J+h) \theta_1 -15J \theta_2 \nonumber \\
\frac{1}{4} \frac{\partial E}{\partial \theta_2} & \approx & -15J \theta_1 + 3 ( 5J - h) \theta_2
\end{eqnarray}

The energy minimum is given by the equation
\begin{eqnarray}
h(h-2J)=0
\end{eqnarray}
The magnetic field of the discontinuity is $h_{d2}=2J$. Setting this value equal to $h_{d1}$ found in App. \ref{appendix:classicallowerfieldmagnetizationdiscontinuityhigherJ} gives $\frac{3}{20}(5+\sqrt{5})$, the value of $J$ for which discontinuities $24'$ and $25'$ together emerge.

\end{appendix}

\bibliography{pentakisdodecahedron}

\begin{table*}
\begin{center}
\caption{$J$ values for which magnetization or susceptibility discontinuities appear or disappear at specific values of the magnetic field $h$ over its saturation value $h_{sat}$ (Fig. \ref{fig:lowestenergyconfigurationdiscontinuities}). Magnetization discontinuities are characterized by a number, and susceptibility discontinuities by a primed number. $N_M$ and $N_{\chi}$ give the total number of magnetization and susceptibility discontinuities right after the $J$ value for which discontinuities appear or disappear. The saturation magnetic field is $h_{sat}=\frac{1}{2}[3+\sqrt{5}+8J+\sqrt{(3+\sqrt{5})^2-4(3+\sqrt{5})J+8(3-\sqrt{5})J^2}]$ for $J \leq \frac{3}{20}(5+\sqrt{5})$, and $8J$ otherwise (App. \ref{appendix:saturationmagneticfield}). The field for the susceptibility discontinuity $24'$ is $h_{d1}=\frac{1}{2}[3+\sqrt{5}+2J-\sqrt{(3+\sqrt{5})^2-16(3+\sqrt{5})J+4(11+2\sqrt{5})J^2}]$ (App. \ref{appendix:classicallowerfieldmagnetizationdiscontinuityhigherJ}), while for susceptibility discontinuity $25'$ it is $h_{d2}=2J$ (App. \ref{appendix:classicalhigherfieldmagnetizationdiscontinuityhigherJ}).}
\vspace{5pt}
\begin{tabular}{c|c|c|c|c||c|c|c|c|c}
 appear & disappear & $J$ & $\frac{h}{h_{sat}}$ & $N_M,N_{\chi}$ & appear & disappear & $J$ & $\frac{h}{h_{sat}}$ & $N_M,N_{\chi}$ \\
\hline
 1 & - & 0 & 0+ & 4,0 & 34 & - & 0.755654 & 0 & 6,3 \\
\hline
 2 & - & 0 & 0.26350 & 4,0 & 35 & 33,$11'$ & 1.010 & 0.132 & 6,2 \\
\hline
 3 & - & 0 & 0.26983 & 4,0 & 36,$12'$ & 4 & 1.012 & 0.597 & 6,3 \\
\hline
 4 & - & 0 & 0.73428 & 4,0 & 37 & 3,36 & 1.015 & 0.597 & 5,3 \\
\hline
 $1',2'$ & - & 0+ & 0+ & 4,2 & 38 & 37,$12'$ & 1.019 & 0.598 & 5,2 \\
\hline
 5,6 & - & 0.228 & 0.07 & 6,2 & 39,40 & 38 & 1.023 & 0.600 & 6,2 \\
\hline
 7,8 & - & 0.229 & 0.07 & 8,2 & 41,42 & - & 1.04939 & 0.04 & 8,2 \\
\hline
 9 & 5,7 & 0.281 & 0.0414 & 7,2 & $13'$ & 41 & 1.04942 & 0.02 & 7,3 \\
\hline
 - & 1,6 & 0.406 & 0.225 & 5,2 & $14'$ & $13'$ & 1.049684 & 0 & 7,3 \\
\hline
 10 & $1'$ & 0.417 & 0.240 & 6,1 & 43 & 34,42 & 1.0497 & 0.11 & 6,3 \\
\hline
 11,$3'$ & 8 & 0.4194 & 0.233 & 6,2 & 44 & 35,43 & 1.0509 & 0.142 & 5,3 \\
\hline
 - & 10,11 & 0.41971 & 0.235 & 4,2 & 45 & 28,44 & 1.0534 & 0.159 & 4,3 \\
\hline
 - & $2',3'$ & 0.41979 & 0.235 & 4,0 & 46,47 & 2 & 1.0542 & 0.549 & 5,3 \\
\hline
 $4',5'$ & - & 0.486 & 0.022 & 4,2 & 48,$15'$ & - & 1.06168 & 0.58 & 6,4 \\
\hline
 12 & $4'$ & 0.489 & 0.021 & 5,1 & 49 & 39,47 & 1.0622 & 0.556 & 5,4 \\
\hline
 13,$6'$ & $5'$ & 0.497 & 0.024 & 6,1 & 50 & 40,48 & 1.06266 & 0.654 & 4,4 \\
\hline
 14,$7'$ & $6'$ & 0.503 & 0.024 & 7,1 & 51 & 46,49 & 1.06271 & 0.548 & 3,4 \\
\hline
 15,$8'$ & 12 & 0.512 & 0.0166 & 7,2 & 52 & 51,$15'$ & 1.0664 & 0.506 & 3,3 \\
\hline
 16 & $7'$ & 0.526 & 0.0228 & 8,1 & 53 & 45,$10'$  & 1.0706 & 0.296 & 3,2 \\
\hline
 17,18 & - & 0.526 & 0.026 & 10,1 & 54 & 53,$9'$ & 1.0714 & 0.308 & 3,1 \\
\hline
 19 & 16,17 & 0.526 & 0.024 & 9,1 & 55,$16'$ & 52 & 1.07262 & 0.425 & 3,2 \\
\hline
 20 & $8'$ & 0.527 & 0.01696 & 10,0 & 56,$17'$ & 54 & 1.073852 & 0.371 & 3,3 \\
\hline
 21 & 13,20 & 0.532 & 0.0168 & 9,0 & 57,$18'$ & 56 & 1.073859 & 0.374 & 3,4 \\
\hline
 22 & 14,19 & 0.534 & 0.0171 & 8,0 & - & 55,57 & 1.07385998 & 0.375 & 1,4 \\
\hline
 23 & 21,22 & 0.535 & 0.0164 & 7,0 & 58,59 & 50 & 1.07625 & 0.897 & 2,4 \\
\hline
 24,25 & 18 & 0.535 & 0.053 & 8,0 & 60,$19'$ & 59 & 1.07643 & 0.8991 & 2,5 \\
\hline
 26,$9'$ & 25 & 0.581 & 0.114 & 8,1 & $20',21'$ & $19'$ & 1.07647 & 0.8996 & 2,6 \\
\hline
 $10'$ & 26 & 0.586 & 0.117 & 7,2 & 61,$22',23'$ & - & 1.0768 & 1 & 3,8 \\
\hline
 27,28 & 24 & 0.588 & 0.0127 & 8,2 & 62 & 60,$22'$ & 1.07923 & 0.933 & 3,7 \\
\hline
 29 & 15,23 & 0.590 & 0.0021 & 7,2 & 63 & 62,$23'$ & 1.07959 & 0.937 & 3,6 \\
\hline
 30 & 9 & 0.591550 & 0 & 7,2 & 64 & 61,63 & 1.080146 & 0.9435 & 2,6   \\
\hline
 31 & 29,30 & 0.596 & 0.0007 & 6,2 & - & 58,$20',21'$ & 1.085 & 0.9101 & 1,4 \\
\hline
 32 & 27,31 & 0.600 & 0.0004 & 5,2 & - & 64 & $\frac{3}{20}(5+\sqrt{5})$ & 1 & 0,4 \\
\hline
 - & 32 & 0.603929 & 0 & 4,2 & $24',25'$ & $14',16',17',18'$ & $\frac{3}{20}(5+\sqrt{5})$ & $\frac{1}{4}$ & 0,2 \\
\hline
 33 & - & 0.620646 & 0 & 5,2 & - & $24'$ & $\frac{5+\sqrt{5}}{4}$ & 0 & 0,1 \\
\hline
 $11'$ & - & 0.64075 & 0 & 5,3 \\
\end{tabular}
\label{table:classicalmagnsuscdisc}
\end{center}
\end{table*}

\begin{table}
\begin{center}
\caption{Ground state of Hamiltonian (\ref{eqn:Hamiltonian}) for $s=\frac{1}{2}$ and zero magnetic field. The columns give the $J$-range for which the state has the lowest energy, its total spin $S$, the irreducible representation of the icosahedral $I_h$ symmetry group it belongs to \cite{Altmann94}, and its degeneracy, which is the product of the multiplicity of the irreducible representation and the degeneracy of an $S$-spin multiplet, $2S+1$. The last column gives the symmetry (s) or antisymmetry (a) of the $S^z=0$ component of the $S$ multiplet with respect to spin inversion.}
\begin{tabular}{c|c|c|c|c}
$J$-range & $S$ & Irreducible & Degeneracy & Spin \\
 & & representation & & inversion \\
\hline
$0 \leq J < 0.371$ & 0 & $A_u$ & 1 & s \\
\hline
$0.371 \leq J \leq 0.642 $ & 2 & $A_g$ & 5 & s \\
\hline
$0.642 < J \leq 1.506 $ & 0 & $A_u$ & 1 & s \\
\hline
$1.506 < J \leq 1.542 $ & 1 & $T_{1u}$ & 9 & a \\
\hline
$1.542 < J < 1.609$ & 2 & $H_g$ & 25 & s \\
\hline
$1.609 \leq J < 1.685$ & 3 & $T_{2u}$ & 21 & a \\
\hline
$1.685 \leq J$ & 4 & $A_g$ & 9 & s
\end{tabular}
\label{table:spinonehalfzerofieldgroundstate}
\end{center}
\end{table}

\begin{table*}
\begin{center}
\caption{Low-energy spectrum of Hamiltonian (\ref{eqn:Hamiltonian}) for $s=\frac{1}{2}$ and different values of $J$ in zero magnetic field. The columns give the reduced energy $\frac{E}{30+60J}$, the total spin $S$, the irreducible representation (irrep.) to which the multiplet belongs \cite{Altmann94}, and its multiplicity mult., which equals the multiplicity of the irrep. times the one of the $S$ sector, which is equal to $2S+1$.}
\vspace{-1pt}
\begin{tabular}{c|c|c|c||c|c|c|c||c|c|c|c||c|c|c|c||c|c|c|c}
$\frac{E}{30+60J}$ & $S$ & irrep. & mult. & $\frac{E}{30+60J}$ & $S$ & irrep. & mult. & $\frac{E}{30+60J}$ & $S$ & irrep. & mult. & $\frac{E}{30+60J}$ & $S$ & irrep. & mult. & $\frac{E}{30+60J}$ & $S$ & irrep. & mult. \\
\hline
\multicolumn{4}{c||} {$J=0$} & \multicolumn{4}{c||} {$J=0.2$} & \multicolumn{4}{c||} {$J=0.3$} & \multicolumn{4}{c||} {$J=0.4$} & \multicolumn{4}{c} {$J=0.5$} \\
\hline
-0.32407 & 0 & $A_u$ & 1 & -0.23675 & 0 & $A_u$ & 1 & -0.21384 & 0 & $A_u$ & 1 & -0.19929 & 2 & $A_g$ & 5 & -0.19068 & 2 & $A_g$ & 5 \\
\hline
-0.31355 & 0 & $H_g$ & 5 & -0.23643 & 1 & $T_{2g}$ & 9 & -0.21332 & 1 & $T_{2g}$ & 9 & -0.19891 & 0 & $A_u$ & 1 & -0.18985 & 2 & $F_g$ & 20 \\
\hline
-0.31175 & 0 & $A_g$ & 1 & -0.23636 & 1 & $T_{1g}$ & 9 & -0.21323 & 1 & $T_{1g}$ & 9 & -0.19875 & 0 & $A_g$ & 1 & -0.18976 & 0 & $A_g$ & 1 \\
\hline
-0.30695 & 1 & $T_{2g}$ & 9 & -0.23619 & 1 & $T_{2u}$ & 9 & -0.21298 & 1 & $T_{2u}$ & 9 & -0.19866 & 1 & $T_{1u}$ & 9 & -0.18950 & 1 & $T_{1u}$ & 9 \\
\hline
-0.30622 & 1 & $F_u$ & 12 & -0.23618 & 0 & $H_u$ & 5 & -0.21288 & 0 & $A_g$ & 1 & -0.19859 & 2 & $F_g$ & 20 & -0.18950 & 2 & $H_u$ & 25 \\
\hline
-0.30435 & 1 & $T_{2u}$ & 9 & -0.23617 & 0 & $F_u$ & 4 & -0.21286 & 0 & $F_u$ & 4 & -0.19849 & 3 & $F_g$ & 28 & -0.18941 & 0 & $A_u$ & 1 \\
\hline
-0.29904 & 1 & $T_{1g}$ & 9 & -0.23613 & 0 & $A_u$ & 1 & -0.21280 & 1 & $H_u$ & 15 & -0.19845 & 2 & $H_u$ & 25 & -0.18932 & 3 & $F_g$ & 28 \\
\hline
-0.29599 & 0 & $H_u$ & 5 & -0.23610 & 0 & $A_g$ & 1 & -0.21276 & 0 & $H_g$ & 5 & -0.19844 & 1 & $T_{1g}$ & 9 & -0.18927 & 1 & $T_{2u}$ & 9 \\
\hline
-0.28983 & 1 & $T_{2u}$ & 9 & -0.23609 & 1 & $H_u$ & 15 & -0.21276 & 0 & $A_u$ & 1 & -0.19844 & 1 & $T_{2u}$ & 9 & -0.18906 & 2 & $H_g$ & 25 \\
\hline
-0.28981 & 1 & $H_u$ & 15 & -0.23608 & 1 & $F_u$ & 12 & -0.21275 & 0 & $H_u$ & 5 & -0.19841 & 0 & $H_g$ & 5 & -0.18905 & 0 & $H_g$ & 5 \\
\hline
-0.28886 & 1 & $T_{1u}$ & 9 & -0.23607 & 0 & $H_u$ & 5 & -0.21272 & 1 & $F_u$ & 12 & -0.19837 & 1 & $T_{2g}$ & 9 & -0.18901 & 1 & $F_u$ & 12 \\
\hline
-0.28834 & 2 & $F_g$ & 20 & -0.23603 & 2 & $H_u$ & 25 & -0.21270 & 2 & $A_g$ & 5 & -0.19820 & 3 & $T_{1u}$ & 21 & -0.18900 & 1 & $T_{2g}$ & 9 \\
\hline
\multicolumn{4}{c||} {$J=0.6$} & \multicolumn{4}{c||} {$J=0.7$} & \multicolumn{4}{c||} {$J=0.8$} & \multicolumn{4}{c||} {$J=0.9$} & \multicolumn{4}{c} {$J=1$} \\
\hline
-0.18517 & 2 & $A_g$ & 5 & -0.18307 & 0 & $A_u$ & 1 & -0.18390 & 0 & $A_u$ & 1 & -0.18566 & 0 & $A_u$ & 1 & -0.18774 & 0 & $A_u$ & 1 \\
\hline
-0.18436 & 0 & $A_u$ & 1 & -0.18216 & 1 & $T_{2g}$ & 9 & -0.18157 & 1 & $T_{2g}$ & 9 & -0.18243 & 0 & $F_g$ & 4 & -0.18458 & 0 & $F_g$ & 4 \\
\hline
-0.18423 & 2 & $F_g$ & 20 & -0.18168 & 1 & $T_{1u}$ & 9 & -0.18101 & 1 & $T_{1u}$ & 9 & -0.18193 & 0 & $A_g$ & 1 & -0.18383 & 0 & $A_g$ & 1 \\
\hline
-0.18421 & 1 & $T_{2g}$ & 9 & -0.18152 & 2 & $A_g$ & 5 & -0.18075 & 0 & $A_g$ & 1 & -0.18172 & 1 & $T_{2g}$ & 9 & -0.18282 & 0 & $H_g$ & 5 \\
\hline
-0.18415 & 0 & $A_g$ & 1 & -0.18127 & 0 & $A_g$ & 1 & -0.18075 & 0 & $F_g$ & 4 & -0.18111 & 1 & $T_{1u}$ & 9 & -0.18231 & 1 & $T_{2g}$ & 9 \\
\hline
-0.18392 & 1 & $T_{1u}$ & 9 & -0.18118 & 1 & $T_{2u}$ & 9 & -0.18041 & 1 & $T_{2u}$ & 9 & -0.18096 & 0 & $H_u$ & 5 & -0.18225 & 0 & $H_u$ & 5 \\
\hline
-0.18379 & 2 & $H_u$ & 25 & -0.18095 & 1 & $F_g$ & 12 & -0.18040 & 0 & $H_u$ & 5 & -0.18082 & 0 & $H_g$ & 5 & -0.18196 & 0 & $H_g$ & 5 \\
\hline
-0.18356 & 1 & $F_g$ & 12 & -0.18095 & 0 & $H_u$ & 5 & -0.18008 & 1 & $F_g$ & 12 & -0.18046 & 1 & $T_{2u}$ & 9 & -0.18165 & 1 & $T_{1u}$ & 9 \\
\hline
-0.18355 & 1 & $T_{2u}$ & 9 & -0.18055 & 2 & $F_g$ & 20 & -0.17990 & 0 & $H_g$ & 5 & -0.18044 & 0 & $H_g$ & 5 & -0.18160 & 1 & $A_u$ & 3 \\
\hline
-0.18349 & 1 & $T_{1u}$ & 9 & -0.18054 & 1 & $F_u$ & 12 & -0.17948 & 1 & $F_u$ & 12 & -0.18016 & 1 & $F_g$ & 12 & -0.18106 & 1 & $F_g$ & 12 \\
\hline
-0.18340 & 1 & $F_u$ & 12 & -0.18051 & 0 & $H_g$ & 5 & -0.17947 & 0 & $H_g$ & 5 & -0.17941 & 1 & $A_u$ & 3 & -0.18105 & 0 & $F_u$ & 4 \\
\hline
-0.18339 & 1 & $T_{2u}$ & 9 & -0.18043 & 0 & $F_g$ & 4 & -0.17925 & 1 & $T_{1g}$ & 9 & -0.17940 & 1 & $F_u$ & 12 & -0.18091 & 1 & $T_{2u}$ & 9 \\
\hline
\multicolumn{4}{c||} {$J=1.1$} & \multicolumn{4}{c||} {$J=1.2$} & \multicolumn{4}{c||} {$J=1.3$} & \multicolumn{4}{c||} {$J=1.4$} & \multicolumn{4}{c} {$J=1.5$} \\
\hline
-0.18985 & 0 & $A_u$ & 1 & -0.19190 & 0 & $A_u$ & 1 & -0.19385 & 0 & $A_u$ & 1 & -0.19568 & 0 & $A_u$ & 1 & -0.19739 & 0 & $A_u$ & 1 \\
\hline
-0.18680 & 0 & $F_g$ & 4 & -0.18895 & 0 & $F_g$ & 4 & -0.19116 & 1 & $T_{1u}$ & 9 & -0.19437 & 1 & $T_{1u}$ & 9 & -0.19732 & 1 & $T_{1u}$ & 9 \\
\hline
-0.18595 & 0 & $A_g$ & 1 & -0.18807 & 0 & $A_g$ & 1 & -0.19100 & 0 & $F_g$ & 4 & -0.19408 & 1 & $T_{2g}$ & 9 & -0.19702 & 1 & $T_{2g}$ & 9 \\
\hline
-0.18504 & 0 & $H_g$ & 5 & -0.18767 & 1 & $T_{1u}$ & 9 & -0.19089 & 1 & $T_{2g}$ & 9 & -0.19292 & 0 & $F_g$ & 4 & -0.19679 & 2 & $H_g$ & 25 \\
\hline
-0.18405 & 0 & $H_u$ & 5 & -0.18745 & 1 & $T_{2g}$ & 9 & -0.19014 & 0 & $A_g$ & 1 & -0.19264 & 1 & $F_u$ & 12 & -0.19649 & 2 & $A_g$ & 5 \\
\hline
-0.18403 & 1 & $T_{2g}$ & 9 & -0.18725 & 0 & $H_g$ & 5 & -0.18936 & 1 & $F_u$ & 12 & -0.19246 & 2 & $H_g$ & 25 & -0.19574 & 2 & $H_u$ & 25 \\
\hline
-0.18397 & 1 & $T_{1u}$ & 9 & -0.18611 & 0 & $H_g$ & 5 & -0.18936 & 0 & $H_g$ & 5 & -0.19220 & 0 & $A_g$ & 1 & -0.19566 & 1 & $F_u$ & 12 \\
\hline
-0.18395 & 0 & $H_g$ & 5 & -0.18610 & 0 & $H_u$ & 5 & -0.18860 & 1 & $F_g$ & 12 & -0.19214 & 2 & $A_g$ & 5 & -0.19538 & 3 & $T_{2u}$ & 21 \\
\hline
-0.18375 & 1 & $A_u$ & 3 & -0.18582 & 1 & $F_u$ & 12 & -0.18831 & 0 & $H_g$ & 5 & -0.19181 & 1 & $F_g$ & 12 & -0.19481 & 1 & $F_g$ & 12 \\
\hline
-0.18343 & 0 & $F_u$ & 4 & -0.18578 & 1 & $A_u$ & 3 & -0.18822 & 0 & $H_u$ & 5 & -0.19145 & 1 & $T_{2u}$ & 9 & -0.19472 & 0 & $F_g$ & 4 \\
\hline
-0.18305 & 0 & $T_{2u}$ & 3 & -0.18571 & 0 & $F_u$ & 4 & -0.18814 & 1 & $T_{2u}$ & 9 & -0.19139 & 2 & $H_u$ & 25 & -0.19452 & 1 & $T_{2u}$ & 9 \\
\hline
-0.18294 & 1 & $F_g$ & 12 & -0.18541 & 1 & $F_g$ & 12 & -0.18785 & 0 & $F_u$ & 4 & -0.19135 & 0 & $H_g$ & 5 & -0.19442 & 0 & $A_g$ & 1 \\
\hline
\multicolumn{4}{c||} {$J=1.6$} & \multicolumn{4}{c||} {$J=1.7$} & \multicolumn{4}{c||} {$J=1.8$} & \multicolumn{4}{c||} {$J=2$} & \multicolumn{4}{c} {$J \to \infty$} \\
\hline
-0.20073 & 2 & $H_g$ & 25 & -0.20558 & 4 & $A_g$ & 9 & -0.21106 & 4 & $A_g$ & 9 & -0.22074 & 4 & $A_g$ & 9 & -0.33288 & 4 & $A_g$ & 9 \\
\hline
-0.20062 & 3 & $T_{2u}$ & 21 & -0.20540 & 3 & $T_{2u}$ & 21 & -0.20977 & 3 & $T_{2u}$ & 21 & -0.21749 & 3 & $T_{2u}$ & 21 & -0.31975 & 3 & $T_{1u}$ & 21 \\
\hline
-0.20045 & 2 & $A_g$ & 5 & -0.20433 & 2 & $H_g$ & 25 & -0.20819 & 3 & $F_g$ & 28 & -0.21599 & 3 & $F_g$ & 28 & -0.30980 & 3 & $H_g$ & 35 \\
\hline
-0.20003 & 1 & $T_{1u}$ & 9 & -0.20407 & 2 & $A_g$ & 5 & -0.20764 & 2 & $H_g$ & 25 & -0.21371 & 3 & $T_{1u}$ & 21 & -0.30756 & 3 & $T_{2u}$ & 21 \\
\hline
-0.19973 & 1 & $T_{2g}$ & 9 & -0.20377 & 3 & $F_g$ & 28 & -0.20738 & 2 & $A_g$ & 5 & -0.21350 & 2 & $H_g$ & 25 & -0.30708 & 2 & $H_g$ & 25 \\
\hline
-0.19970 & 2 & $H_u$ & 25 & -0.20333 & 2 & $H_u$ & 25 & -0.20666 & 2 & $H_u$ & 25 & -0.21325 & 2 & $A_g$ & 5 & -0.30681 & 3 & $F_g$ & 28 \\
\hline
-0.19958 & 4 & $A_g$ & 9 & -0.20253 & 1 & $T_{1u}$ & 9 & -0.20485 & 1 & $T_{1u}$ & 9 & -0.21255 & 2 & $H_u$ & 25 & -0.30656 & 5 & $A_g$ & 11 \\
\hline
-0.19900 & 0 & $A_u$ & 1 & -0.20222 & 1 & $T_{2g}$ & 9 & -0.20462 & 3 & $T_{1u}$ & 21 & -0.21098 & 2 & $H_g$ & 25 & -0.30604 & 3 & $F_u$ & 28 \\
\hline
-0.19895 & 3 & $F_g$ & 28 & -0.20107 & 2 & $F_g$ & 20 & -0.20451 & 1 & $T_{2g}$ & 9 & -0.21045 & 2 & $F_g$ & 20 & -0.30356 & 2 & $A_g$ & 5 \\
\hline
-0.19843 & 1 & $F_u$ & 12 & -0.20099 & 1 & $F_u$ & 12 & -0.20445 & 2 & $F_g$ & 20 & -0.20994 & 2 & $F_u$ & 20 & -0.29817 & 4 & $T_{1u}$ & 27 \\
\hline
-0.19756 & 1 & $F_g$ & 12 & -0.20051 & 0 & $A_u$ & 1 & -0.20379 & 2 & $H_g$ & 25 & -0.20951 & 3 & $H_g$ & 35 & -0.29802 & 2 & $T_{2u}$ & 15 \\
\hline
-0.19738 & 2 & $F_g$ & 20 & -0.20032 & 2 & $F_u$ & 20 & -0.20376 & 2 & $F_u$ & 20 & -0.20926 & 3 & $F_u$ & 28 & -0.29726 & 2 & $H_u$ & 25
\end{tabular}
\label{table:lowenergyspectrumspinonehalf}
\end{center}
\end{table*}

\begin{table}
\begin{center}
\caption{Ground-state magnetization discontinuities with $\Delta S^z > 1$ for Hamiltonian (\ref{eqn:Hamiltonian}) when $s=\frac{1}{2}$. The columns give the $J$-range of the magnetization discontinuity, the $S^z$ value below and above the discontinuity, and the irreducible representation (irrep.) of the icosahedral $I_h$ symmetry group the lowest-energy states below and above the discontinuity belong to \cite{Altmann94}.}
\begin{tabular}{c|c|c|c|c}
$J$-range & $S^z_{below}$ & $S^z_{above}$ & Irrep. below & Irrep. above \\
\hline
$0 \leq J \leq 1.012$ & 10 & 12 & $A_g$ & $A_u$ \\
\hline
$0.279 < J \leq 0.302$ & 1 & 3 & $T_{2g}$ & $F_g$ \\
\hline
$0.302 < J < 0.307$ & 0 & 3 & $A_u$ & $F_g$ \\
\hline
$0.307 \leq J \leq 0.371 $ & 0 & 2 & $A_u$ & $A_g$ \\
\hline
$0.642 < J < 0.743 $ & 0 & 2 & $A_u$ & $A_g$ \\
\hline
$0.707 < J \leq 1.032$ & 4 & 6 & $A_g$ & $A_u$ \\
\hline
$0.980 \leq J \leq 1.071$ & 6 & 8 & $A_u$ & $A_g$ \\
\hline
$1.050 \leq J \leq 1.056 $ & 4 & 6 & $A_g$ & $A_u$ \\
\hline
$1.074 < J \leq 1.075 $ & 6 & 8 & $A_g$ & $A_g$
\end{tabular}
\label{table:spinonehalfmagnetizationdiscontinuities}
\end{center}
\end{table}

%

\end{document}